\documentclass[review,onecolumn,times,authoryear]{elsarticle}

\usepackage{graphicx}
\usepackage{amssymb}
\usepackage{natbib}
\usepackage{amsmath}

\journal{Icarus}
\newcommand\icarus{Icarus\ }
\newcommand\apj{Astrophys. J.\ }

\newcommand\nat{Nature\ }
\newcommand\mnras{Mon.\ Not.\ R.\ Astron.\ Soc.\ }
\def\memsai{\ref@jnl{Mem.~Soc.~Astron.~Italiana}}

\makeatletter
\let\jnl@style=\rm
\def\ref@jnl#1{{\jnl@style#1}}
\def\araa{\ref@jnl{ARA\&A}}             
\def\nar{\ref@jnl{New A Rev.}}          
\def\aj{\ref@jnl{AJ}}                   
\makeatother

\def\figurecont#1{%
    \let\oldthefigure=\thefigure
    \def\thefigure{#1 (Cont.)}
    \gdef\figurecontlabel{#1}
}

\def\endfigurecont{%
    \let\thefigure=\oldthefigure
    \def\setc{\setcounter{figure}}
    \expandafter\expandafter\expandafter\setc\expandafter{\figurecontlabel}
}

\usepackage{bm}
\usepackage{epstopdf}
\usepackage{pgfplots}

\def\ten#1{\bm{#1}}
\def\differ{{\rm d}}
\let\vec\boldsymbol

\def\Tr{{\rm Tr\,}}

\long\def\cb#1\ce{}

\begin{document}

\begin{frontmatter}

\title{SPH/$N$-body simulations of small ($D = 10\,{\rm km}$) asteroidal breakups
and improved parametric relations for Monte--Carlo collisional models}

\author{P.~\v Seve\v cek$^{\rm a,\star}$}
\author{M.~Bro\v z$^{\rm a}$}
\author{D.~Nesvorn\'y$^{\rm b}$}
\author{B.~Enke$^{\rm b}$}
\author{D.~Durda$^{\rm b}$}
\author{K.~Walsh$^{\rm b}$}
\author{D.~C.~Richardson$^{\rm d}$}
    
\address{$^{\rm a}$ Institute of Astronomy, Charles University, Prague, V Hole\v sovi\v ck\'ach 2, 18000 Prague 8, Czech Republic}
\address{$^{\rm b}$ Southwest Research Institute, 1050 Walnut Street, Suite 400, Boulder, CO 80302, USA}
\address{$^{\rm c}$ Earth Sciences Department, University of California Santa Cruz, Santa Cruz, CA 95064, USA}
\address{$^{\rm d}$ Department of Astronomy, University of Maryland, College Park, MD 20742, USA}    


\begin{abstract}
    We report on our study of asteroidal breakups, i.e.~fragmentations of targets,  
    subsequent gravitational reaccumulation and formation of small asteroid families. 
    We focused on parent bodies with diameters $D_{\rm pb} = 10\,\rm km$. 
    Simulations were performed with a smoothed-particle hydrodynamics (SPH) code combined with 
    an efficient $N$-body integrator. 
    We assumed various projectile sizes,
    impact velocities and impact angles (125 runs in total).
    Resulting size-frequency distributions are significantly different from scaled-down 
    simulations with $D_{\rm pb} = 100\,\rm km$ targets \citep{Durda_etal_2007Icar..186..498D}.
    We derive new parametric relations describing fragment distributions,
    suitable for Monte-Carlo collisional models.
    We also characterize velocity fields and angular distributions of fragments, which
    can be used as initial conditions for $N$-body simulations of small asteroid families.
    Finally, we discuss a number of uncertainties related to SPH simulations.

\end{abstract}


\end{frontmatter}




\renewcommand{\thefootnote}{\fnsymbol{footnote}}

\section{Introduction and motivation}\label{sec:introduction}

Collisions between asteroids play an important role in the evolution 
of the main belt. Understanding the fragmentation process
and subsequent reaccumulation of fragments is
crucial for studies of the formation of the solar system or 
the internal structure of the asteroids.
Remnants of past break-ups are 
preserved to a certain extent in the form of asteroid families -- 
groups of asteroids located close to each other in the space of proper
elements $a_{\rm p}$, $e_{\rm p}$, $I_{\rm p}$
\citep{Hirayama_1918, Nesvorny_etal_2015arXiv150201628N}.

The observed size-frequency distribution (SFD) of the family members  
contains a lot of information and can aid us to determine
the mass $M_{\rm PB}$ of the parent body.
However, it cannot be determined by merely summing up the observed  
family members, as a large portion of the total mass is presumably
'hidden' in fragments well under observational completeness.
The SFD is also modified over time, due to ongoing secondary collisional evolution and dynamical removal by the Yarkovsky drift and various gravitational resonances, etc.
This makes the procedure a bit difficult for ancient asteroid families 
and relatively simple for very young ($<10\,\rm Myr$) clusters, such as 
Karin or Veritas \citep{Nesvorny_etal_2006Icar..183..296N, Michel_etal_2011Icar..211..535M}.

Disruptive and cratering impacts have been studied experimentally, 
using impacts into cement mortar targets 
\citep[e.g.][]{Davis_Ryan_1990, Nakamura_Fujiwara_1991}.
However, in order to compare those results to impacts of asteroids
we need to scale the results up in terms of the mass of the target
and kinetic energy of the projectile by several orders of magnitude.
The scaled impact experiments can still have significantly different outcomes,
compared to the asteroid collisions, due to the increasing role of 
gravitational compression, different fragmentation mechanisms etc.
Experiments yield valuable information about properties of materials, but they 
are not sufficient to unambiguously determine results of asteroid collisions.

Numerical simulations are thus used to solve a standard set of 
hydrodynamic equations; however, the physics of fragmentation is much more 
complex than that.
Especially for low-energy cratering impacts, 
it is necessary to simulate an explicit propagation of cracks in the target.
There is no \emph{ab initio} theory of fragmentation,
but phenomenological theories has been developed to describe
the fragmentation process, 
such as the Grady--Kipp model of fragmentation
\citep{Grady_Kipp_1980}, used in this paper, 
or more complex models including porosity based on the P-$\alpha$ model
\citep{Herrmann_1969}.


Common methods of choice for studying impacts are 
shock-physics codes and particle codes \citep{Jutzi_2015}.
The most important outputs of simulations are masses $M_{\rm lr}$ and $M_{\rm lf}$ of 
the largest remnant and largest fragment, respectively, and the exponent $q$ of the power-law
approximation to the cumulative size-frequency distribution $N(>$$D)$,
i.\,e.\,the number $N$ of family members with diameter larger than given~$D$.
Parametric relations, describing the dependence of $M_{\rm lr}$ and $q$
on input parameters, can be then applied on collisional models of the main asteroid belt,
such as those presented in \cite{Morbidelli_etal_2009Icar..204..558M} or \cite{Cibulkova_etal_2014Icar..241..358C}; 
however, if we aim to determine the size of the parent body,
%
%
we need to
solve an \emph{inverse} problem. 

A single simulation gives us the SFD for a given
size of the parent body and several parameters of the impactor.
However, if one wishes to derive the size of the parent body and 
impactor parameters from the observed SFD, 
it is necessary to conduct a large set of simulations with 
different parameters
and then find the SFD that resembles the observed one as accurately as possible.
This makes the problem difficult as the parameter space is quite extensive. 
For one run, we usually have to specify
the parent body size $D_{\rm PB}$,
the projectile size $d_{\rm project}$,
the impact speed $v_{\rm imp}$,
and the impact angle $\phi_{\rm imp}$ 
(i.e.~the angle between the velocity vector of the impactor
and the inward normal of the target at the point of collision).
Other parameters of the problem are the material properties of 
considered asteroids, such as bulk density, shear modulus, porosity etc.

Due to the extent of the parameter space, a thorough study would be highly demanding on computational resources. It is therefore reasonable to fix the size of the parent body and study breakups 
with various parameters of the impactor.

A large set of simulations was published by \cite{Durda_etal_2007Icar..186..498D},
who studied disruptions of 100\,km monolithic targets.
Similarly, 
\cite{Benavidez_etal_2012Icar..219...57B} performed an analogous 
set of simulations with rubble-pile targets.
They also used the resulting SFDs to estimate the size of the parent body 
for a number of asteroid families.
As the diameter of the parent body is never exactly 100\,km,
the computed SFDs have to be multiplied by a suitable scaling factor $f_{\rm scale}$
to match the observed one. 
However, small families have been already discovered (e.g.~Datura, \cite{Nesvorny_etal_2015arXiv150201628N})
and their parent-body size is likely $D_{\rm pb} = 10\,\rm km$, i.e.~an order-of-magnitude smaller.
The linearity of the scaling is a crucial assumption
and we will assess the plausibility of this assumption in this paper.

To fill up a gap in the parameter space,
we proceed with small targets.
We carried out a set of simulations with $D_{\rm pb} = 10\,\rm km$ parent bodies 
and carefully compared them with the simulations of \cite{Durda_etal_2007Icar..186..498D}.

The paper is organised as follows.
In Section \ref{sec:numerical_methods}, we briefly describe our numerical methods.
The results of simulations are presented in Section 3. 
Using the computed SFDs we derive parametric relations for 
the slope $q$ and the masses $M_{\rm lr}$ and $M_{\rm lf}$ of the largest remnant 
and the largest fragment, respectively, in Section 4.
Finally, we summarize our work in Section 5.


\section{Numerical methods}
\label{sec:numerical_methods}
We follow a hybrid approach of \cite{Michel_2001,Michel_2002,Michel_2003,Michel_2004},
employing
an SPH discretization for the simulation of fragmentation 
and an $N$-body integrator for subsequent gravitational reaccumulation.
Each simulation can be thus divided into three phases:
i)~a fragmentation,
ii)~a hand-off, and
iii)~a reaccumulation.
We shall describe them sequentially in the following subsections.


\subsection{Fragmentation phase}

The first phase of the collision is described by hydrodynamical equations
in a lagrangian frame. They properly account for supersonic shock wave
propagation and fragmentation of the material.
We use the \texttt{SPH5} code by \cite{Benz_Asphaug_1994}
for their numerical solution.
In the following, we present only a brief description of 
equations used in our simulations and we refer readers
to extensive reviews of the method \citep{Rosswog_2009, Cossins_2010, Price_2008, Price_2012} for a more detailed description.

Our problem is specified by four basic equations, namely
the equation of continuity,
equation of motion,
energy equation
and Hooke's law:
\begin{eqnarray}
{\differ\rho\over\differ t} &=& -\rho\nabla\cdot\vec v\,, \label{rce_kontinuity_lagrange} \\
{\differ\vec v\over\differ t} &=& {1\over\rho}\nabla\cdot\ten\sigma \,,\label{Navier_Stokes_lagrange} \\
    {\differ U\over\differ t} &=&  -{P\over\rho}\Tr\dot{\ten\epsilon} + {1\over\rho}\ten S : \dot{\ten\epsilon} \,,\label{prvni_veta_td_lagrange} \\
    {\differ\ten S\over\differ t} &=& 2\mu\left(\dot{\ten\epsilon} - {\textstyle{1\over 3}}\ten 1\,\Tr\dot{\ten\epsilon}\,\right) \,,\label{rce_konstitucni}
\end{eqnarray}
supplemented by the Tillotson equation of state \citep{Tillotson_1962}.
The notation is as follows:
$\rho$~is the density,
$\vec v$~the speed,
$\ten\sigma$~the stress tensor (total), where $\ten\sigma \equiv -P\ten 1 + \ten S$,
$P$~the pressure,
$\ten 1$~the unit tensor,
$\ten S$~the deviatoric stress tensor,
$U$~the specific internal energy,
$\dot{\ten\epsilon}$ the strain rate tensor, where $\dot{\ten\epsilon} \equiv {\textstyle{1\over 2}}\left[\nabla\vec v + (\nabla\vec v)^{\rm T}\right]$,
with its trace $\Tr\dot{\ten\epsilon} = \nabla\cdot\vec v$,
$\mu$~the shear modulus.

The model includes both elastic and plastic deformation,
namely the yielding criterion of \cite{vonMises_1913} ---
given by the factor $f \equiv \min[Y_0^2/({3\over 2}\ten S:\ten S), 1]$ ---
and also failure of the material. The initial distribution of cracks
and their growth to fractures is described by models
of \cite{Weibull_1939} and \cite{Grady_Kipp_1980}, which use a scalar
parameter~${\cal D} \in \langle0, 1\rangle$ called damage,
as explained in \cite{Benz_Asphaug_1994}.
The stress tensor of damaged material is then modified as
$\ten\sigma = - (1-{\cal D}H(-P))P\ten 1 + (1-{\cal D})f\ten S$,
where $H(x)$ denotes the Heaviside step function.
In this phase, we neglect the influence of gravity, 
which is a major simplification of the problem.

In a smoothed-particle hydrodynamic (SPH) formalism,
Eqs.~(\ref{rce_kontinuity_lagrange}) to (\ref{rce_konstitucni})
are rewritten so as to describe an evolution of individual SPH particles
(denoted by the index $i = 1..N$):
\begin{eqnarray}
    {\differ\rho_i\over\differ t} &=& -\rho_i\sum_j \frac{m_j}{\rho_j}(\vec v_j-\vec v_i) \cdot \nabla W_{ij} \,,\label{rce_kontinuity_sph}\\
    {\differ\vec v_i\over\differ t} &=& \sum_j m_j \left(\frac{\ten\sigma_i + \ten\sigma_j}{\rho_i\rho_j} \right)\cdot\nabla W_{ij}\,,\label{navier_stokes_sph}\\
{\differ U_i\over\differ t} &=& -{P_i\over\rho_i}\sum_\gamma \dot\epsilon_i^{\gamma\gamma} + {1\over\rho_i}\sum_\alpha\sum_\beta S_i^{\alpha\beta} \dot\epsilon_i^{\alpha\beta}
    \,,\label{prvni_veta_td_sph}\\
    {\differ\ten S_i\over\differ t} &=& 2\mu\left(\dot{\ten\epsilon}_i - {\textstyle{1\over3}}\ten 1\sum_\gamma\dot\epsilon_i^{\gamma\gamma}\right) \,,\label{rce_konstitucni_sph}
\end{eqnarray}
with: 
\begin{equation}
\dot\epsilon_i^{\alpha\beta} = {1\over2\rho_i} \sum_j m_j \left[ (v_j^\alpha-v_i^\alpha){\partial W_{ij}\over\partial x^\beta} + (v_j^\beta-v_i^\beta){\partial W_{ij}\over\partial x^\alpha} \right] \,,\label{dot_epsilon_sph}
\end{equation}
where
$m_j$ denote the masses of the individual SPH particles,
$W_{ij} \equiv W(|\vec r_i-\vec r_j|, h)$ the kernel function,
$h$~the symmetrized smoothing length, $h = {1\over 2}(h_i+h_j)$.
Both the equation of motion and the energy equation were also supplied
with the standard artificial viscosity term $\Pi_{ij}$
 \citep{Monaghan_Gingold_1983}:
 \begin{equation}
    \Pi_{ij} = \left\{\begin{array}{ll} 
        \frac{1}{\rho}\left(-\alpha c_{\rm s} \mu_{ij} + \beta \mu_{ij}^2\right) & (\vec v_i-\vec v_j)\cdot (\vec r_i-\vec r_j) \leq 0\,,\\
            0 & \rm otherwise \,,
                    \end{array}\right.
\end{equation}
where:
\begin{equation}
    \mu_{ij} = \frac{h(\vec v_i-\vec v_j)\cdot (\vec r_i-\vec r_j)}
    {\|\vec r_i-\vec r_j\|^2 + \epsilon h^2 }\,,
\end{equation}
$c_{\rm s}$ is the sound speed, $\alpha_{\rm AV} = 1.5$ and $\beta_{\rm AV} = 3$.
We sum over all particles, but since the kernel has a compact support, 
the algorithm has an asymptotic complexity ${\cal O}(N N_{\rm neighbours})$.
The actual number of SPH particles we used is $N \doteq 1.4\times 10^5$,
and the number of neighbours is usually $N_{\rm neighbours} \simeq 50$. 
There is also an evolution equation for the
smoothing length~$h_i$ in order to adapt to varying distances
between SPH particles.


\subsection{Hand-off procedure}
Although SPH is a versatile method suitable for simulating 
both the fragmentation and the gravitational reaccumulation, 
the time step of the method is bounded by the Courant criterion 
and the required number of time steps for complete reaccumulation
is prohibitive.
In order to proceed with inevitably simplified but efficient computations,
we have to convert SPH particles to solid spheres, a procedure called hand-off.
In this paper, we compute the corresponding radius~$R_i$ as:
\begin{equation}
R_i = \left({3m_i\over4\pi\rho_i}\right)^{1\over3}\,.\label{handoff}
\end{equation}

The time $t_{\rm handoff}$ at which the hand-off takes place is determined
by three conditions:
\begin{enumerate}
    \item It has to be at least $2D_{\rm PB}/c_{\rm s} \simeq 1\,{\rm s}$ ($c_{\rm s}$ being the sound speed), 
i.e. until the shock wave and rarefaction wave propagate across the target;
 \item Fractures (damage) in the target should not propagate anymore,
even though in catastrophic disruptions the shock wave usually damages
the whole target and material is then practically strengthless;
 \item The pressure in the fragmented parent body should be zero so that the corresponding
acceleration $-{1\over\rho}\nabla P$ is zero, or at least negligible.
According to our tests for $D_{\rm PB} = 10\,{\rm km}$ targets,
such relaxation takes up to $10\,{\rm s}$.
\end{enumerate}
On the other hand, there is an upper limit for $t_{\rm handoff}$ given
by the gravitational acceleration of the target, $g = GM_{\rm PB}/R_{\rm PB}^2$,
which has to be small compared to the escape speed~$v_{\rm esc} = \sqrt{2GM_{\rm PB}/R_{\rm PB}}$,
i.e. a typical ejection speed $v_{\rm ej}$ of fragments.
The corresponding time span should thus be definitely shorter than
$v_{\rm esc}/g \simeq 10^3\,{\rm s}$.


\subsection{Reaccumulation phase}

Finally, gravitational reaccumulation of now spherical fragments
is computed with an $N$-body approach. We use the \texttt{pkdgrav} code
as modified by \cite{Richardson_etal_2000Icar..143...45R} for this purpose. 
It accounts for mutual gravitational interactions between fragments:
\begin{equation}
\ddot{\vec r_i} = -\sum_{j\ne i} {Gm_j\over r_{ij}^3}\vec r_{ij}\,,\label{pkdgrav}
\end{equation}
An~${\cal O}(N^2)$ problem is simplified significantly using a tree code algorithm, 
i.e.~by clustering fragments to cells and evaluating gravitational moments up to hexadecapole order,
provided they fit within the opening angle $\differ\theta = 0.5\,{\rm rad}$.
The time step was $\Delta t = 10^{-6}$ (in $G = 1$ units, or about $5\,{\rm s}$ in SI),
and the time span $50,000\,\Delta t$, long enough that the reaccumulation
is over, or negligible.

Regarding mutual collisions, we assumed perfect sticking only,
meaning no bouncing or friction.
Consequently, we have no information about resulting shapes of fragments,
we rather focus on their sizes, velocities and corresponding statistics.


\section{A grid of simulations for $D_{\rm PB} = 10\,{\rm km}$ targets}
We performed a number of simulations with 
$D_{\rm pb} = 10\,\rm km$ parent bodies,
impact speed $v_{\rm imp}$ varying from 3 to $7 \,\rm km/s$,
diameter $d_{\rm project}$ of the impactor from $0.293\,\rm km$
to $1.848\,\rm km$ (with a logarithmic stepping) and
the impact angle $\phi_{\rm imp}$ from $15^\circ$ to $75^\circ$.
The kinetic energy of the impact:
\begin{equation}
    Q = \frac{\frac{1}{2}m_{\rm project} v_{\rm imp}^2}{M_{\rm pb}}
\end{equation}
therefore varies from $\sim 10^{-2} Q_{\rm D}^\star$ 
to $\sim 20 Q_{\rm D}^\star$, where $Q_{\rm D}^\star$ is 
the critical energy for shattering and dispersing 50\%
of the parent body. We adopted $Q_{\rm D}^\star(D)$ values for 
comparisons from the scaling law of \cite{Benz_Asphaug_1999Icar..142....5B}.
The total number of performed runs is~125.
We assume a monolithic structure of both the target and the impactor, 
and the material properties were selected those of basalt
(summarized in Table~\ref{tab:constants}).

\begin{table}[t]
    \centering
\begin{tabular}{ll}
    \multicolumn{2}{c}{Material parameters}\\
    \hline\hline
    density at zero pressure & $\rho = 2700\,\rm kg/m^3$ \\
    bulk modulus & $A = 2.67\cdot 10^{10} \,\rm Pa$ \\
    non-linear Tillotson term & $B =  2.67 \cdot 10^{10}\,\rm Pa$ \\
    sublimation energy & $u_0 = 4.87 \cdot 10^8\,\rm J/kg$ \\ 
    energy of incipient vaporization & $u_{\rm iv} = 4.72\cdot 10^6\,\rm  J/kg$ \\
    energy of complete vaporization  & $u_{\rm cv} = 1.82 \cdot 10^7\,\rm J/kg$ \\
    shear modulus & $\mu = 2.27\cdot 10^{10}\,\rm Pa$ \\
    von Mises elasticity limit & $Y_0=  3.50\cdot 10^{9} \,\rm Pa$\\
    Weibull coefficient & $k = 4.00\cdot 10^{29}$ \\
    Weibull exponent & $m = 9$ \\
    \hline\\
    \multicolumn{2}{c}{SPH parameters} \\
       \hline\hline
    number of particles in target & $ N_{\rm pb} = 10^5$ \\
    number of particles in projectile & $ N_{\rm pb} = 100 \mbox{ to } 630$ \\
    Courant number & $C=1$ \\
    linear term of artificial viscosity & $\alpha_{\rm AV} = 1.5$ \\
    quadratic term of artificial viscosity & $\beta_{\rm AV} = 3.0$\\
    duration of fragmentation phase& $t_{\rm handoff} = 10\,\rm s$
\end{tabular}
    \caption{Constant parameters used in our SPH simulations.}
    \label{tab:constants}
\end{table}

\subsection{Size-frequency distributions}
\label{sec:SFD}

For each run we constructed a cumulative size-frequency distributions 
$N(>$$D)$ of fragments and we plotted them in
Fig.~1.


At first sight, the SFDs are well-behaved. 
Both cratering and catastrophic events
produce mostly power-law-like distributions.
Some distributions, mainly those around $Q/Q_{\rm D}^\star \sim 1$, 
have an increasing slope at small sizes (at around $D\sim 0.3\,\rm km$),
but since this is close to the resolution limit, 
it is possibly a numerical artifact. 

For supercatastrophic impacts with $d_{\rm project} = 1.848\,\rm km$, 
the distributions differ from power laws substantially;
the slope becomes much steeper at large sizes of fragments.
These are the cases where the gap between 
the largest remnant and the largest fragment
disappears (we therefore say the largest remnant does not exist).

The situation is quite different for impacts with an oblique impact angle, 
mainly for $\phi_{\rm imp} = 75^\circ$. We notice that these impacts 
appear much less energetic compared to other impact angles, even
though the ratio $Q/Q_{\rm D}^\star$ is the same.
The cause of this apparent discrepancy is simply the geometry of the impact.
At high impact angles
, the impactor does not hit the target 
with all its cross-section and a part of it misses the target entirely
\citep[grazing impacts, see][]{Leinhardt_Stewart_2012ApJ...745...79L}.
Therefore, a part of the kinetic energy is not deposited
into the target and the impact appears less energetic,
compared to head-on impacts.


\begin{figure*}
\centering
\hbox{%
    \hskip-70pt%
    \includegraphics[width=17.5cm]{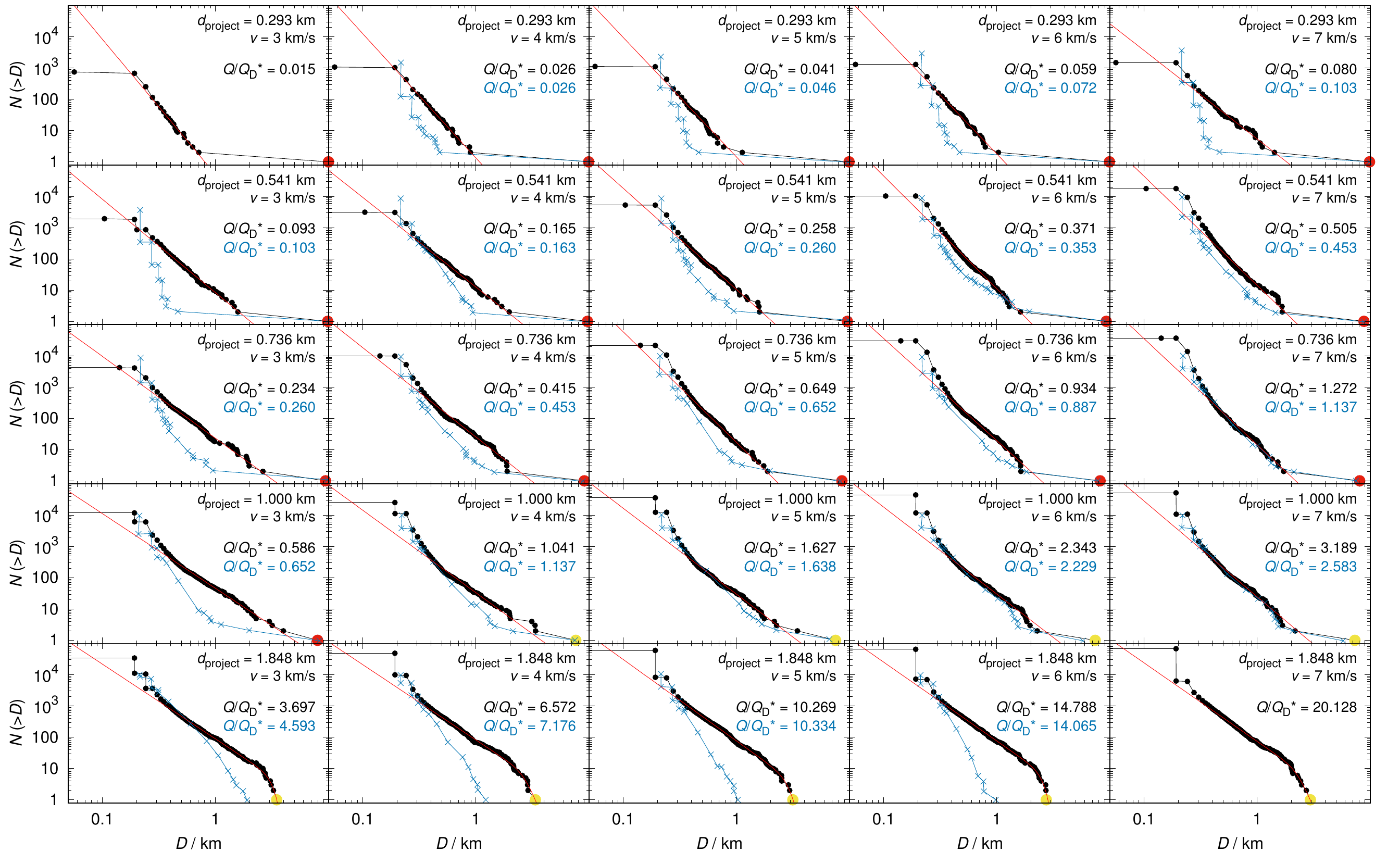}
    }
\caption{Cumulative size-frequency distributions $N({>}D)$ of fragments
ejected during disruptions of parent bodies with sizes $D_{\rm PB} = 10\,{\rm km}$.
The impact angle was $\phi_{\rm imp} = 45^\circ$; results for different impact angles are shown in Appendix D.
The projectile size is increasing downwards, from 
    $d_{\rm project} = 0.293\,\rm km$ to $1.848\,\rm km$,
so that the logarithm of the mass ratio $\log_{10} (m_{\rm project}/M_{\rm PB}) = 3.0, 2.6, 2.2, 1.8$ and $1.0$.
The impact speed is increasing to the right, from $v_{\rm imp} = 3$ to $7\,{\rm km}\,{\rm s}^{-1}$.
Both of the quantities are also indicated in individual panels,
together with the ratio $Q/Q^\star_{\rm D}$ of the specific energy~$Q$
and strength~$Q^\star_{\rm D}$ inferred from the scaling law of \cite{Benz_Asphaug_1999Icar..142....5B}.
Largest remnant size $D_{\rm LR}$ is coloured red or yellow for cratering 
or catastrophic events, respectively.
For a discussion of scaling we overplot simulated SFD's from \cite{Durda_etal_2007Icar..186..498D}
computed for disruptions of $D_{\rm PB} = 100\,{\rm km}$ targets and scaled
down by dividing sizes by a factor of 10 (blue lines and labels).
To compare `apples with apples', we compare runs with
(approximately) the same $Q/Q^\star_{\rm D}$ ratios and the same impact angle.
    For some impact parameters, the scaled SFD is missing as there is no run in the dataset of \cite{Durda_etal_2007Icar..186..498D} with comparable $Q/Q_{\rm D}^\star$.
Finally, the red curves are fits of a suitable function, used to derive 
    parametric relations (see Section \ref{sec:parametric}).
}
\label{GRID_1km_size_distribution_SCALED_100km}
\end{figure*}

\subsection{Speed histograms}

\begin{figure*}
\centering
\hbox{%
    \hskip-70pt%
    \includegraphics[width=18cm]{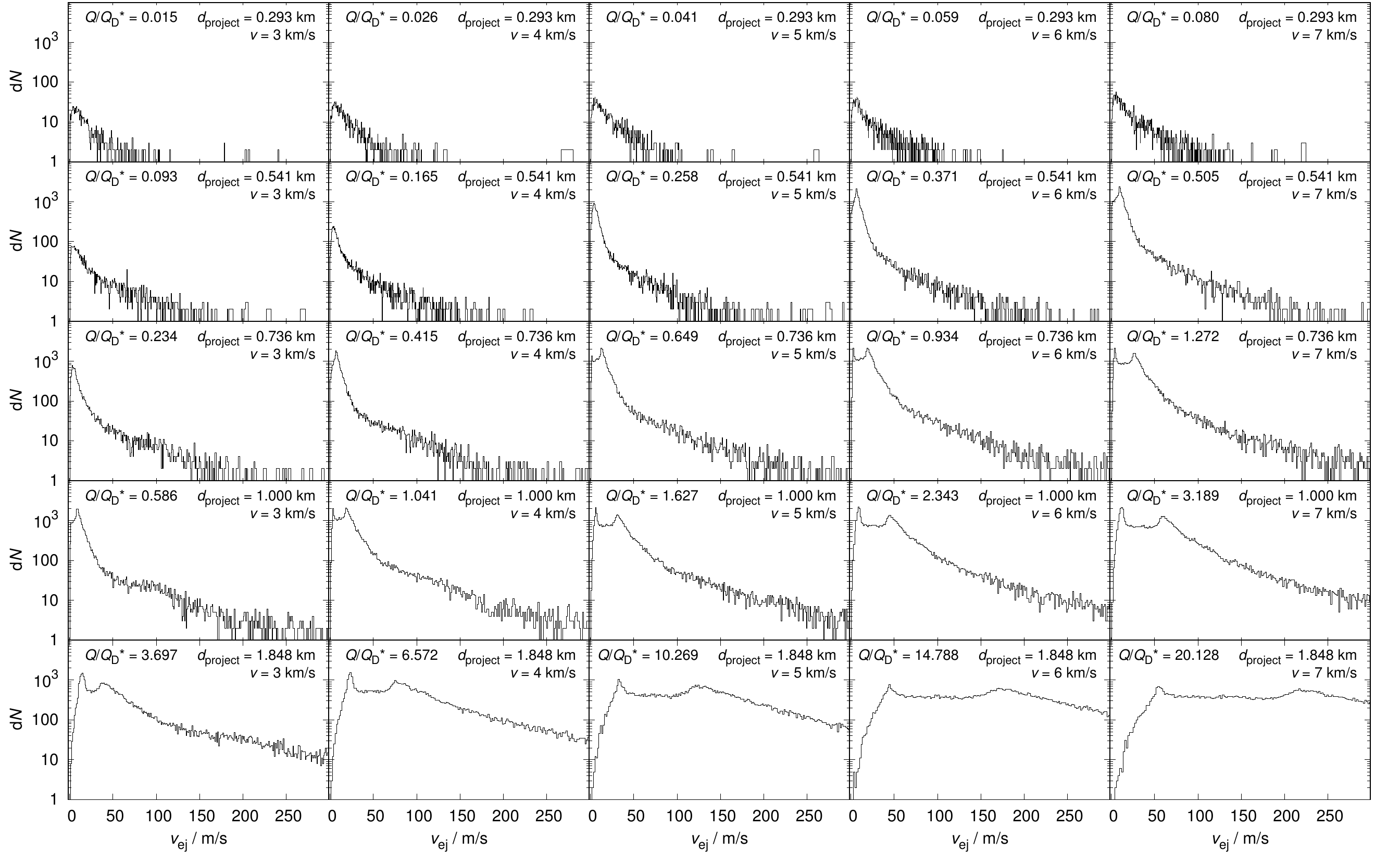}
    }
\caption{Differential histograms ${\rm d}N$ of ejection speeds~$v_{\rm ej}$ of fragments
for the same set of simulations as in Fig.~\ref{GRID_1km_size_distribution_SCALED_100km}.
The speed is computed in a barycentric reference frame
    with outliers ($v_{\rm imp}>1\,\rm km/s$) are removed 
    as they are mostly remnants of the projectile.    
The escape speed from the target $D_{\rm PB} = 10\,{\rm km}$ in size
is $v_{\rm esc} = 6.1\,{\rm m}\,{\rm s}^{-1}$, histogram peaks are thus of order $v_{\rm esc}$, at least for the majority of simulations.
However, there is also a significant second peak
visible. It is close to the first peak 
    for cratering to mid-energy impacts and
    extends to speeds $v_{\rm ej} > 100\,\rm m/s$
for supercatastrophic breakups with $Q/Q^\star_{\rm D} \gtrsim 10$. 
    The impact angle $\phi_{\rm imp}=45^\circ$ in this case.}
\label{GRID_1km_hist_velocity}
\end{figure*}

Similarly to the size-frequency distributions, we computed
speed distributions of fragments.
The results are shown in Fig.~\ref{GRID_1km_hist_velocity}.
As we are computing an absolute value of the velocity,
the resulting histogram depends on a selected reference frame.
We chose a barycentric system for all simulations;
however, we excluded high-speed remainders
of the projectile with velocities $v_{\rm ej }> v_{\rm cut} \equiv 1\,\rm km/s$. 
These outliers naturally
appear mainly for oblique impact angles.
Because of very large ejection velocities, such fragments
cannot belong to observed families and 
if we had included them in the constructed velocity field of the synthetic family,
it would artificially shift velocities of fragments to higher values.

The main feature of cratering events is the peak
around the escape velocity $v_{\rm esc}$. 
This  peak is created by
fragments ejected at the point of impact.
With an increasing impact energy, the tail of the histogram 
extends as the fragments are ejected at higher velocities.

Interestingly, there is a second peak 
at around $Q/Q_{\rm D}^\star \sim 0.3$.
This is because of ejection of fragments from the antipode of the target.
If the shockwave is energetic enough,
it causes an ejection of many fragments.
The second  peak is barely visible at oblique 
impact angles.

\subsection{Isotropy vs anisotropy of the velocity field}
\begin{figure*}
\centering
\hbox{%
    \hskip-70pt%
\includegraphics[width=18cm]{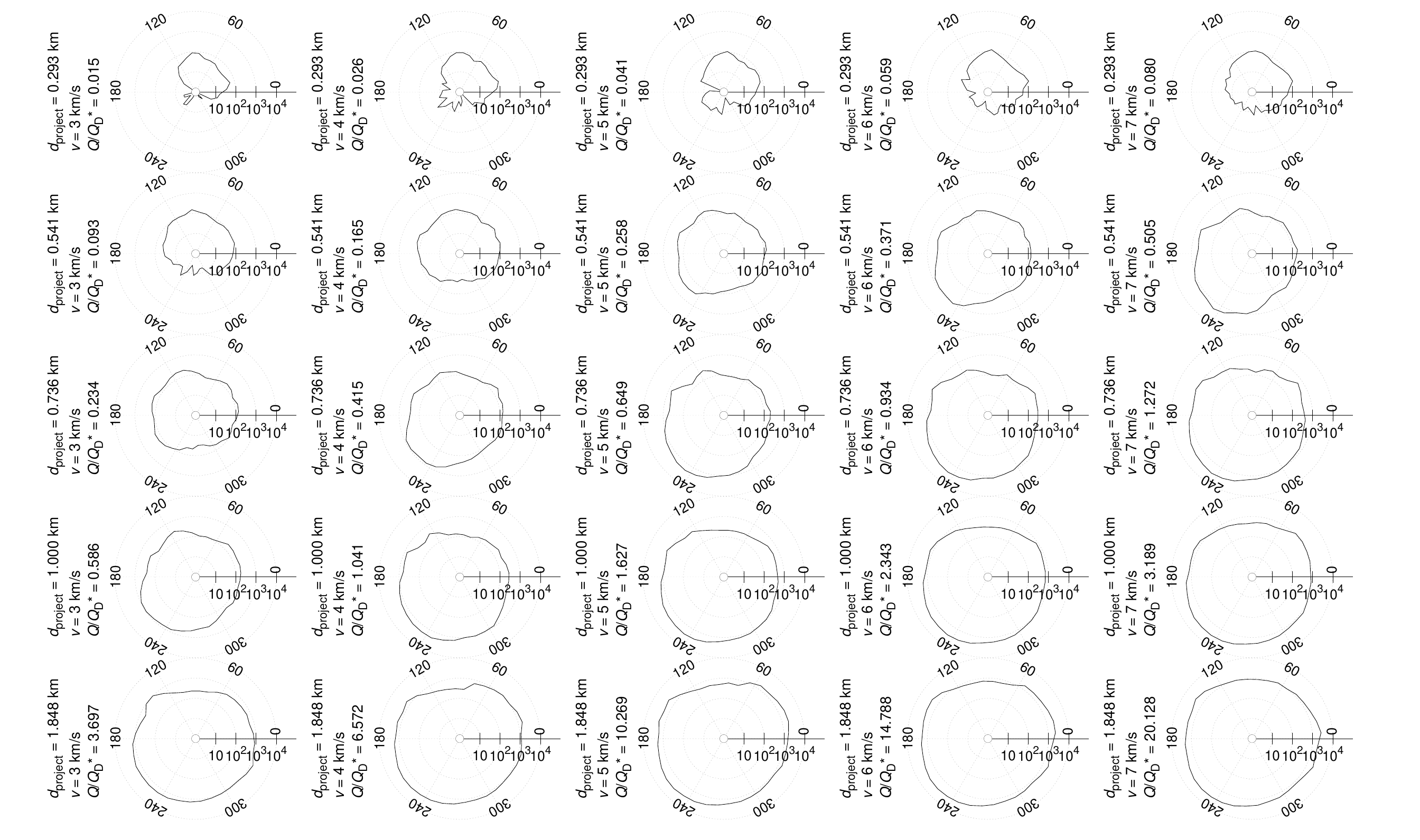}
    }
\caption{Histograms of velocity angular distribution (in the plane of the collision) of 
fragments. The velocities are evaluated in the barycentric coordinate system with outliers
removed. The angle $180^\circ$ corresponds to the velocity direction of the projectile. 
    The impact angle $\phi_{\rm imp}= 45^\circ$. }
\label{GRID_1km_hist_angular}
\end{figure*}

Fig.~\ref{GRID_1km_hist_angular} shows angular distributions
of the velocity fields in the plane of the impact. The histograms are drawn
as polar plots with a $5^\circ$ binning. The angles on plots 
correspond to the points of impact for given impact angle $\phi_{\rm imp}$;
for cratering events, all the ejecta are produced at the point of impact 
and the distribution of fragments is therefore nicely clustered around $\phi_{\rm imp}$.

Cratering impacts tend to produce velocity fields mainly in the direction of the impact angle.
Catastrophic impacts, on the other hand, generally produce much more isotropic velocity fields.
However, the isotropy is not perfect, even though we removed 
outliers as above. Even for the supercatastrophic impacts,
the number of fragments in different directions can vary by a factor of 5.
Further changes of the reference frame may improve the isotropy.
Note that for observed families, it is also not clear
where is the reference points, as the identification of 
family members (and interlopers) is ambiguous.


\subsection{A comparison with scaled-down $\it D_{\rm PB} = 100\,{\it km\/}$ simulations}

The mid-energy events with $Q/Q_{\rm D}^\star \sim 1$ have SFDs 
comparable
to scaled 100\,km ones. In this regime, down-scaling of the distribution
for $D_{\rm pb} = 100\,\rm km$ targets seems to be a justifiable way to approximate SFDs 
for targets of smaller sizes.

In case of cratering events, however, our simulations differ
significantly from scaled ones. Impacts into 10\,km
targets produce a much shallower fragment distribution compared to 100\,km impacts;
see impacts with $d_{\rm project} = 0.293\,\rm km$.
We also note that supercatastrophic runs have different
outcomes than the 100\,km ones; our  distributions
are much shallower and have a much larger largest fragment.
They also have a steeper part of the SFD at larger diameters,
which is \emph{not} visible for 100\,km simulations, 
at least not to the same extent.


\section{Parametric relations for Monte-Carlo collisional models}
\label{sec:parametric}

Size-frequency distributions constructed from our simulations 
mostly have a power-law shape
with a separated largest remnant.
The slope of the distribution in a log-log plot
can be therefore fitted with a linear function:
\begin{equation}
    \log N(>\!D) = q \log [D]_{\rm km} + c\,.
\end{equation}
Supercatastrophic events behave differently though,  
and their SFDs can be well fitted with a two-slope function:
\begin{equation}
    \log N(>\!D) = K\left(\log [D]_{\rm km} - \log [D_0]_{\rm km}\right)+c\,,
\end{equation}
where:
\begin{equation}
    K(x) = \frac{1}{2}(q_1+q_2)x+ \frac{1}{2}\frac{q_1-q_2}{k}\log\left(2\cosh k x\right)\,.
    \label{eq:knee_function}
\end{equation}
In this approximation of the SFD, $q_1$ and $q_2$ are the limit slopes for $D \rightarrow \infty$ and $D \rightarrow 0$, respectively, and $k$ characterizes the 
``bend-off`` of the function.
As the fitting function is highly non-linear and the dependence on $k$ is very weak
(given rather sparse input data), the fit doesn't generally converge,
we thus fix $k=10$ and perform the fit using only four parameters:
$s_1, s_2, D_0$ and~$c$. 


Because impacts at high angles appear weaker due
the geometry (see Section \ref{sec:SFD}),
we have to account for the actual kinetic energy 
delivered into the target. 
We chose a slightly different approach than 
\cite{Leinhardt_Stewart_2012ApJ...745...79L}
and modified the specific impact energy $Q$ by a ratio 
of the cross-sectional area of the impact and the total 
area of the impactor. 
Using a formula for circle-circle intersection:
let $R$ be the radius of the target, $r$ the radius of the projectile
and $d$ a projected distance between their centers.
The area of impact is then given by:
\begin{align}
\nonumber    A &= r^2 \cos^{-1} \left( \frac{d^2 + r^2 - R^2}{2dr} \right)
    + R^2 \cos^{-1} \left(  \frac{d^2 + R^2 - r^2}{2dR} \right)- \\
        &- \frac{1}{2} \sqrt{ (R+r-d)(d+r-R)(d-r+R)(d+r+R) }\,.
    \label{eq: }
\end{align}
As both spheres touch at the point of the impact, we have:
\begin{equation}
    d = (r+R) \sin \phi_{\rm imp}\,.
    \label{eq: }
\end{equation}
Using these auxiliary quantities, 
we define the \emph{effective} specific impact energy:
\begin{equation}
    Q_{\rm eff} = Q \frac{A}{\pi r^2}\,.
    \label{eq:effective_impact_energy}
\end{equation}

\begin{figure*}
\hbox{%
    \hskip-70pt%
    \includegraphics[width=18cm]{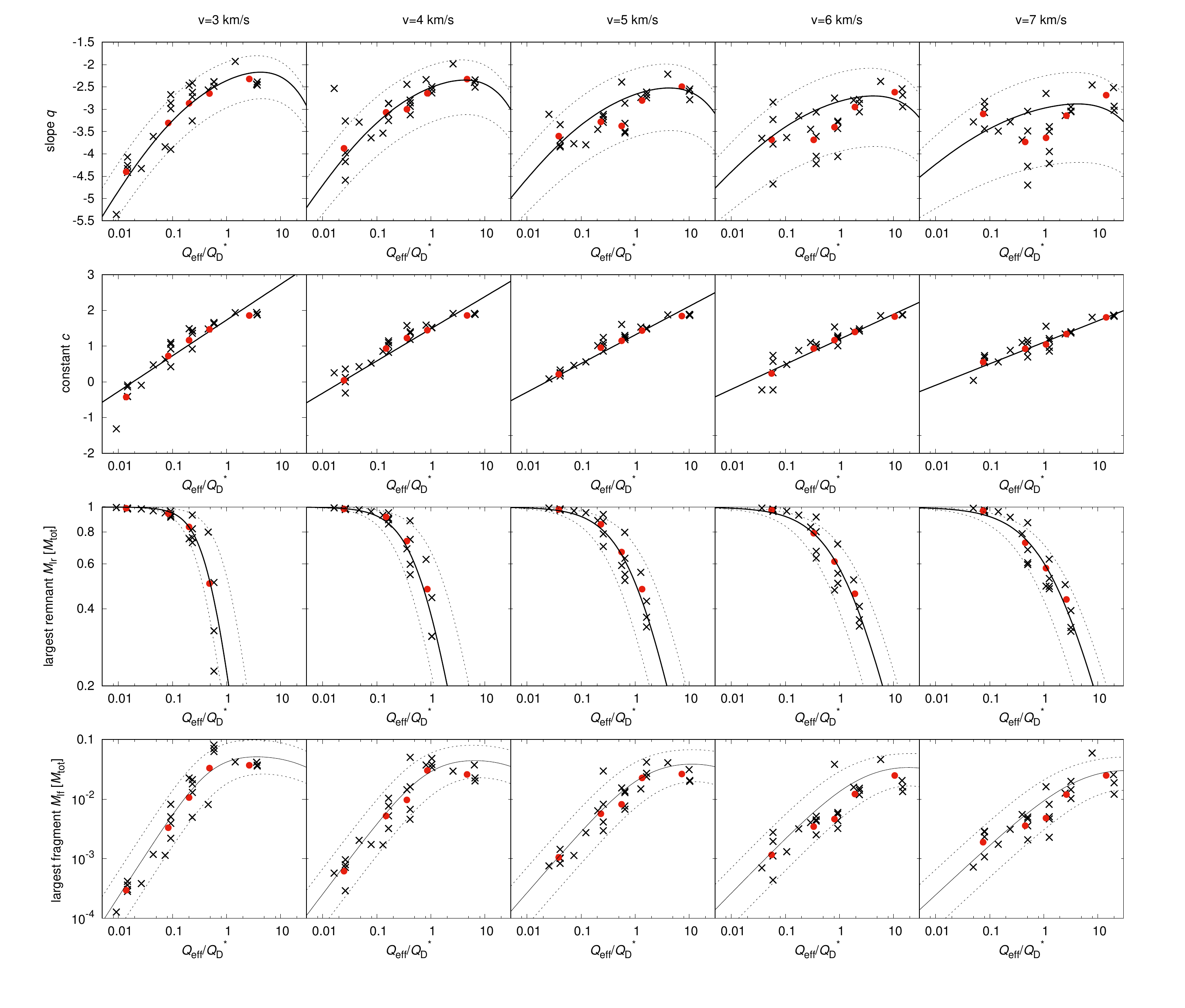}
    }
    \caption{Parameters of the power-law fits of 
    size-frequency distributions
    (first and second row) and masses
     of the largest remnant $M_{\rm lr}$ and the largest fragment 
     $M_{\rm lf}$ (third and fourth row) as functions of the effective impact 
     energy $Q_{\rm eff}/Q_{\rm D}^\star$, defined
     by Eq.\,(\ref{eq:effective_impact_energy}).
     We plotted these quantities for each value of impact speed
     separately as considering $Q_{\rm eff}/Q_{\rm D}^\star$ as a single
     parameter  would imply a large variance of data
     and therefore a large uncertainty of parametric relations.
     Each black cross represents one SPH/$N$-body simulation,
     and the red circles are given by averaging over impact angles 
     $\phi_{\rm imp}$.
     The data are fitted with suitable functions
     and the scatter of values propagates to the parametric
     relations as uncertainties, 
     (see Section \ref{sec:parametric}).
     }
     \label{img:parametric}
\end{figure*}

In Fig.~\ref{img:parametric},
we separately plot slopes $q$, constants $c$ 
of the linear fits of the SFDs, and the masses of the largest remnants 
$M_{\rm lr}$ and largest fragment $M_{\rm lf}$.
Each of these quantities shows a distinct dependence on the impact speed $v_{\rm imp}$,
suggesting parametric relations cannot be well described 
by a single parameter $Q_{\rm eff}/ Q_{\rm D}$.
We therefore plot each dependence separately for 
different $v_{\rm imp}$
and we explicitly express the dependence on 
$v_{\rm imp}$ in parametric relations.

For low speeds, slopes $q$ can be reasonably fitted with a function:
\begin{equation}
    q = -12.3+0.75 v_{\rm imp} 
        + \cfrac{(11.5-1^{+0.2}_{-0.1}v_{\rm imp})\exp\left(-5\cdot 10^{-3} \cfrac{Q_{\rm eff}}{Q_{\rm D}^\star}\right)  }{1+0.1^{+0.01}_{-0.02}
    \left(\cfrac{Q_{\rm eff}}{Q_{\rm D}^\star}\right)^{%
            -0.4 }}\,,
    \label{eq:slope_parametric}
\end{equation}
where $v_{\rm imp}$ is expressed in $\rm km/s$.
However, for high speeds (especially for $v=7\,\rm km/s$),
the individual values of $q$ for different impact angles 
differ significantly and thus the fit has a very high uncertainty.
We account for this behaviour in Eq.\,(\ref{eq:slope_parametric}),
where the uncertainty increases with an increasing speed.

The constant $c$ can be well fitted by linear function:
\begin{equation}
    c = 0.9 + 2.3 \exp(-0.35v_{\rm imp}) + \left(1.3 - 0.1v_{\rm imp}\right) \left(\frac{Q_{\rm eff}}{Q_{\rm D}^\star}\right)\,.
\end{equation}
The high scatter noted in the parametric relation for the slope $q$
is not present here.
This parameter is of lesser importance for Monte-Carlo models though, as 
the distribution must be normalized anyway to conserve the total mass.

Largest remnants are also plotted in Fig.~\ref{img:parametric}.
Notice that some points are missing here as the largest remnant 
does not exist for supercatastrophic impacts. 
As we are using the effective impact energy $Q_{\rm eff}$
as an independent variable, 
the runs with impact angle $\phi=75^\circ$ 
produce largest remnants of sizes comparable 
to other impact angles.
This helps to decrease the scatter 
of points and make 
the derived parametric relation more accurate.
We selected a fitting function:
\begin{equation}
    M_{\rm lr} = \frac{M_{\rm tot}}{1 + \left[0.6^{+0.5}_{-0.2} + 56 \exp(-1.0^{+0.6}_{-0.2}v_{\rm imp})\right] \left(\cfrac{Q_{\rm eff}}{Q_{\rm D}^\star}\right)%
    ^{0.8+8\exp(-0.7v_{\rm imp})}}  \,. 
    \label{eq: }
\end{equation}

Largest fragments (fourth row) exhibit 
a larger scatter, similarly as the slopes $q$.
The masses of the largest fragment can differ by an order of magnitude
for different impact angles (notice the logarithmic scale on the $y$-axis).
Nevertheless, the values averaged over impact angles (red circles)
lie close the fit in most cases.
The fitting function for the largest remnant is:
{\small
\begin{equation}
    M_{\rm lf} = \frac{%
        M_{\rm tot} }{%
            0.24^{+0.60}_{-0.15} v_{\rm imp}^3 \left(\cfrac{Q_{\rm eff}}{Q_{\rm D}^\star}\right)^%
            {-0.6 - 2\exp(-0.3v_{\rm imp})} 
           \kern-40pt + \exp\left(-0.3^{+0.2}_{-0.2}v_{\rm imp}\right) \cfrac{Q_{\rm eff}}{Q_{\rm D}^\star}
            + 11^{+15}_{-8} +2v_{\rm imp}} \,.
\end{equation}
}
This function bends and starts to decrease for $Q/Q_{\rm D}^\star \gg 1$.
Even though this behaviour is not immediately evident from the  plotted points,
the largest fragment \emph{must} become a decreasing 
function of impact energy in the  supercatastrophic
regime.

\section{Conclusions and future work}\label{sec:conclusions}
In this paper, we studied disruptions and subsequent gravitational
reaccumulation of asteroids with diameter $D_{\rm pb}=10\,\rm km$.
Using an SPH code and an efficient $N$-body integrator, 
we performed impact simulations for various projectile sizes $d_{\rm project}$,
impact speeds $v_{\rm imp}$ and angles $\phi_{\rm imp}$.
The size-frequency distributions, constructed from the results of our simulations,
appear similar to the scaled-down simulations of \cite{Durda_etal_2007Icar..186..498D}
only in the transition regime between cratering and catastrophic events ($Q/Q_{\rm D}^\star \simeq 1$);
however, they differ significantly for both the weak cratering impacts and for supercatastrophic impacts.

The resulting size-frequency distributions can be used to estimate the size 
of the parent body, especially for small families. 
As an example, we used our set of simulations to determine $D_{\rm pb}$ of the Karin family.
This cluster was studied in detail by \cite{Nesvorny_etal_2006Icar..183..296N}
and we thus do not intend to increase the accuracy of their result,
but rather to assess the uncertainty of linear SFD scaling.
The closest fit to the observed SFD of the Karin cluster yields a parent body with 
$D_{\rm pb}=25\,\rm km$ --- a smaller, but comparable value to $D_{\rm pb}=33\,\rm km$,
obtained by \cite{Nesvorny_etal_2006Icar..183..296N}. 
Using the set of $D_{\rm pb}=100\,\rm km$ simulations,
\cite{Durda_etal_2007Icar..186..498D} obtained an estimate 
$D_{\rm pb}\simeq 60\,\rm km$. It is therefore reasonable
that the best estimate is intermediate between 
the result from upscaled 10\,km runs and downscaled 100\,km runs.
We do not consider our result based on ``generic'' simulations more accurate
than the result of \cite{Nesvorny_etal_2006Icar..183..296N};
however, the difference between the results can be seen as 
an estimate of uncertainty one can expect when scaling 
the SFDs by a factor of~3.

We derived new parametric relations, describing the masses $M_{\rm lr}$
and $M_{\rm lf}$ of the largest remnant and the largest fragment,
respectively, and the slope $q$ of the size-frequency distribution
as functions of the impact parameters. These parametric relations
can be used straightforwardly to improve the accuracy of collisional models, as
the fragments created by a disruption of small bodies were
previously estimated as scaled-down disruptions of $D_{\rm pb}=100\,\rm km$ bodies.

In our simulations, we always assumed monolithic targets. 
The results can be substantially different for porous bodies, though,  
as the internal friction has a significant influence on the fragmentation
\citep{Jutzi_2015, Asphaug_2015}.
This requires using a different yielding model, such as Drucker--Prager
criterion.
We postpone a detailed comparison between monolithic and porous bodies
for future work.


\section*{Acknowledgements}\label{sec:acknowledgements}
The work of MB and P\v S was supported by the
Grant Agency of the Czech Republic (grant no. P209/15/04816S).


\appendix




\section{Initial distribution of SPH particles}
For a unique solution of evolutionary differential equations, 
initial conditions have to be specified. In our case, this means 
setting the initial positions and velocities of SPH particles.
We assume non-rotating bodies, all particles of the target 
are therefore at rest and all particles of the impactor
move with the speed of the impactor.

Optimal initial positions of SPH particles have to meet several criteria.
First of all, the particles have to be distributed evenly in space.
This requirement eliminates a random distribution as a suitable 
method, for using such a distribution would necessarily lead to 
clusters of particles in some parts of space and a lack
of particles in other parts.

We therefore use a  hexagonal-close-packing lattice in the simulations.
They are easily set up and have an optimal interpolation accuracy.
However, no lattice is \emph{isotropic}, 
so there are always preferred directions in the distribution of SPH particles.
This could potentially lead to numerical artifacts, such as pairing instability
\citep{Herant_1994}.
Also, since the particle concentration is uniform,
 the impact is therefore resolved by only a few SPH particles for small impactors.
We can increase accuracy of cratering impacts by distributing SPH particles
nonuniformly, putting more particles at the point of impact and fewer in more distant
places.

Here we assess the uncertainty introduced by 
using different initial conditions of SPH particles.
A suitable method for generating 
a nonuniform isotropic distribution 
has been described by \cite{Diehl_2012} and \cite{Rosswog_2015}.
Using initial conditions generated by this method, we ran several SPH/$N$-body simulations,
and we compared the results to the simulations with lattice initial conditions. 

\begin{figure*}
    \centering
    \includegraphics[width=\textwidth]{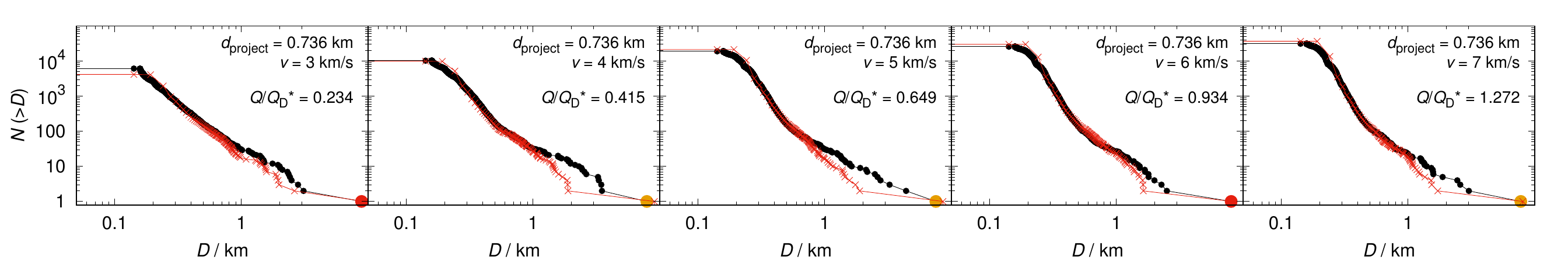}
    \caption{SFDs constructed from five different simulations with $D_{\rm pb}=10\,\rm km$,
    $d_{\rm project} = 0.736\,\rm km$ and impact angle $\phi_{\rm imp}=45^\circ$.
    Black histogram shows the runs with the nonuniform distribution 
    generated by the method of \cite{Diehl_2012}, while red are the previous (lattice) results
    shown in Fig.~\ref{GRID_1km_size_distribution_SCALED_100km}.}
    \label{img:nonuniform}
\end{figure*}

The comparison is in Fig.~\ref{img:nonuniform}.
Generally, the target shatters more for the nonuniform distribution.
The largest remnant is smaller; the difference is up to 10\% for the  performed simulations.
There are also more fragments at larger diameters, compared to the lattice distribution.
This is probably due to slightly worse interpolation properties of the nonuniform
distribution. A test run for a \emph{random} distribution of particles led
to a complete disintegration of the target and a largest remnant smaller
by an order of magnitude, suggesting the smaller largest remnant 
is a numerical artifact of the method.
On the other hand, the SFD is comparable at smaller diameters.
This leads to more bent, less power-law-like SFDs for nonuniform runs.

\cb\subsection{Artificial viscosity}
coefficients $\alpha$, $\beta$:
\begin{equation}
    \Pi_{ij} = \left\{\begin{array}{ll}
            \frac{1}{\rho}(-\alpha\mu_{ij}+\beta\mu_{ij}^2)&(\vec v_i-\vec v_j)\cdot(\vec r_i-\vec r_j)<0 \\
            0 & \rm otherwise 
    \end{array}\right.
    \label{eq: }
\end{equation}
Figure \ref{img:av}.
\begin{figure}
    \centering
    \includegraphics{artificial_viscosity.eps}
    \caption{Size-frequency distribution for different values of coefficients $\alpha$ 
        and $\beta$
    ($\beta=2\alpha$). Those are high-energy head-on impacts
    ($d_{\rm project} =0.074\,\rm km$, $v=7\,\rm km/s$, $\phi=15^\circ$)
    as we believe the influence of artificial viscosity is more  
    significant, compared to cratering events.}
    \label{img:av}
\end{figure}
\ce

\section{Energy conservation vs. timestepping}

\begin{figure*}
\hbox{%
    \hskip-70pt%
\includegraphics[width=18cm]{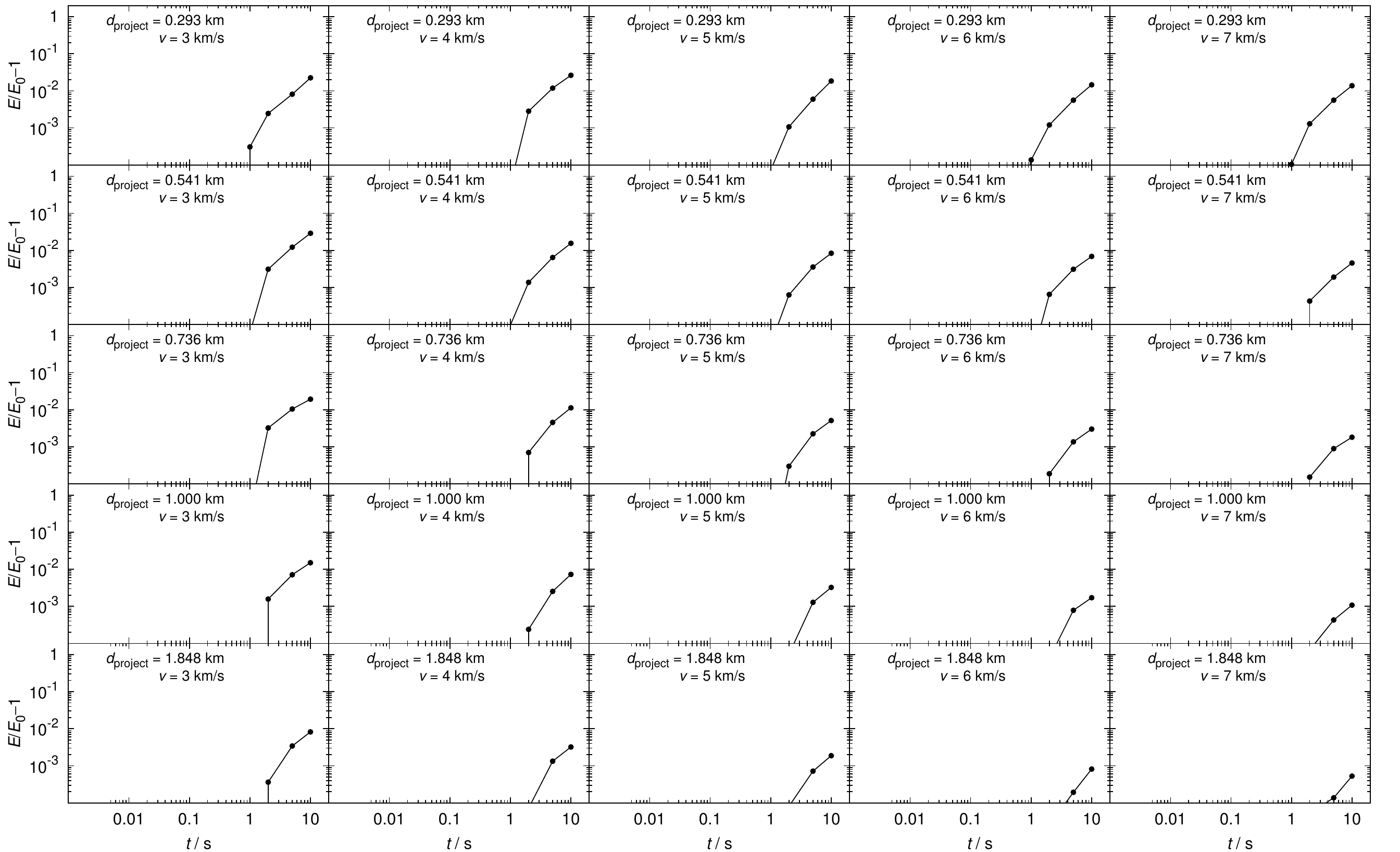}
    }
\caption{Relative total energy~$E$ vs time~$t$ for the same grid
of simulations as presented in Sec.~3.
The diameter of the target was always $D = 10\,{\rm km}$
and the impact angle $\phi_{\rm imp} = 45^\circ$.
The maximum relative energy error is of the order of $10^{-2}$
at the final time $t = 10\,{\rm s}$.}
\label{GRID_10km_10SEC_energy}
\end{figure*}

\begin{figure*}
        \includegraphics[width=8.5cm]{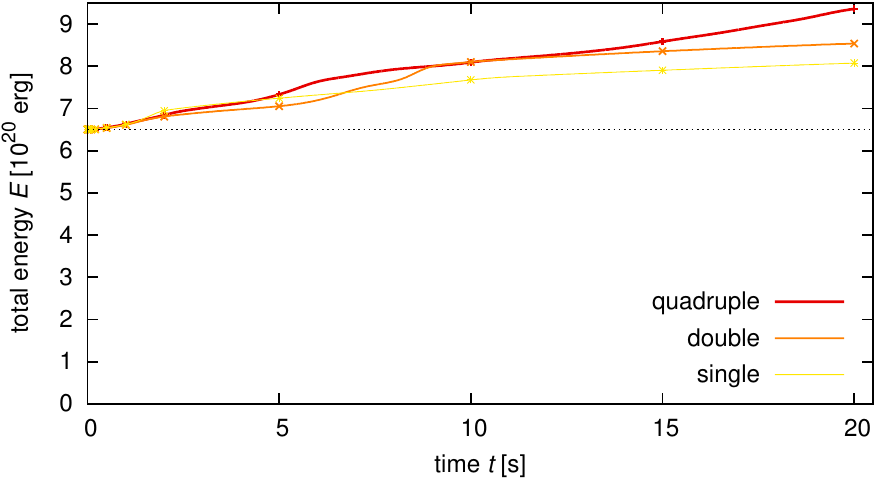}
\caption{Total energy~$E$ vs time~$t$ for a small cratering impact,
with a $D = 1\,{\rm km}$ target and $d = 22\,{\rm m}$ projectile,
and several compiled numerical precisions of the SPH5 code:
quadruple (128-bit), double (64-bit) and single (32-bit).
Because the computation in quadruple precision is very slow,
we use $N \doteq 1.4\times 10^4$ particles only in this test.
Other parameters were set up similarly as in other simulations presented in Sec.~3.
Neither version conserves the energy sufficiently,
which is an indication that round-off errors are {\em not\/}
the dominant cause of the energy increase.}
\label{sph5_code_QUADRUPLE_energy}
\end{figure*}
        \begin{figure*}
\includegraphics[width=8.5cm]{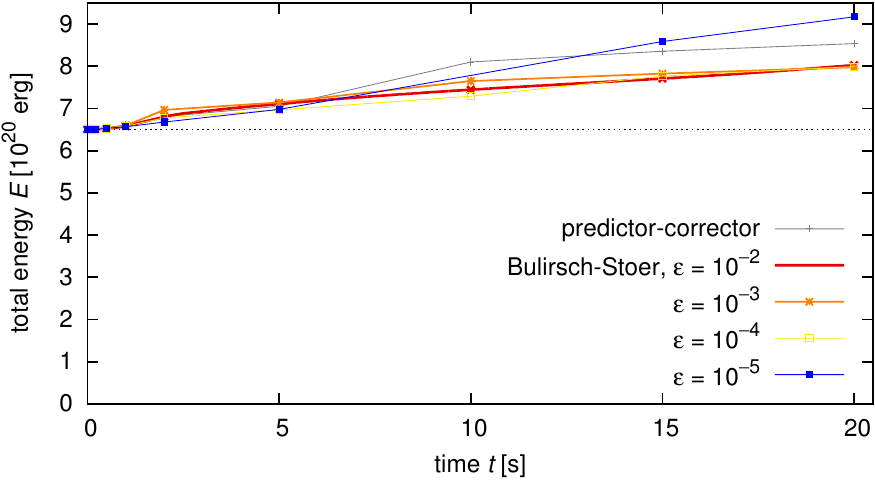}
\caption{Total energy~$E$ vs time~$t$ for the same impact
as in Fig.~\ref{sph5_code_BS_energy}, but with different timestepping schemes,
namely the default predictor--corrector (controlled by the Courant number $C = 1.0$ and other time step restrictions)
and the Bulirch--Stoer with the unitless parameter $\epsilon_{\rm BS} = 10^{-2}$ to $10^{-5}$. 
The scaling by maximum values of variables was used,
which corresponds to constant absolute errors.
The energy conservation was somewhat improved this way,
with the exception of the lowest $\epsilon_{\rm BS}$ at late times.}
\label{sph5_code_BS_energy}
\end{figure*}

Modelling of smaller breakups seems more difficult.
Apart from poor resolution of the impactor,
if one uses the same (optimum) SPH particle mass as in the target,
and a relatively low number of ejected fragments,
weak impacts may also exhibit problems with energy conservation
(see~Fig.~\ref{GRID_10km_10SEC_energy}).
This is even more pronounced in the case of low-speed collisions,
e.g. of $D = 1\,{\rm km}$ target, $d = 22\,{\rm m}$ projectile,
at $v_{\rm imp} = 3\,{\rm km}/{\rm s}$ and $\phi_{\rm imp} = 45^\circ$.

At first, we thought that small oscillations of density
--- with relative changes $\Delta\rho/\rho$ smaller than the numerical precision ---
are poorly resolved, and subsequently cause the total energy to increase.
But when we performed the same simulation in quadruple precision
(with approximately 32 valid digits) we realised there is essentially
no improvement (see Fig.~\ref{sph5_code_QUADRUPLE_energy}),
so this cannot be the true reason.

Instead, we changed the timestepping scheme and superseded
the default predictor/corrector
with the Bulirsch--Stoer integrator \citep{Press_etal_1992nrfa.book.....P},
which performs a series of trial steps with $\Delta t$ divided by factors $2, 4, 6, \dots$,
and checks if the relative difference between successive divisions is less than small dimensionless factor~$\epsilon_{\rm BS}$
and then extrapolates to $\Delta t\to 0$.
In our case, a scaling of quantities is crucial.
In principle, we have three options:
  (i)~scaling by expected maximum values, which results in a constant absolute error;
 (ii)~current values, or constant relative error;
(iii)~derivatives times time step, a.k.a. constant cumulative error.
The option (i) seems the only viable one, otherwise the integrator
is exceedingly slow during the initial pressure build-up.
According to Fig.~\ref{sph5_code_BS_energy},
we have managed to somewhat improve the energy conservation this way,
but more work is needed to resolve this issue.

\section{Energy conservation vs sub-resolution acoustic waves}

\begin{figure}
\centering
\includegraphics[width=8.5cm]{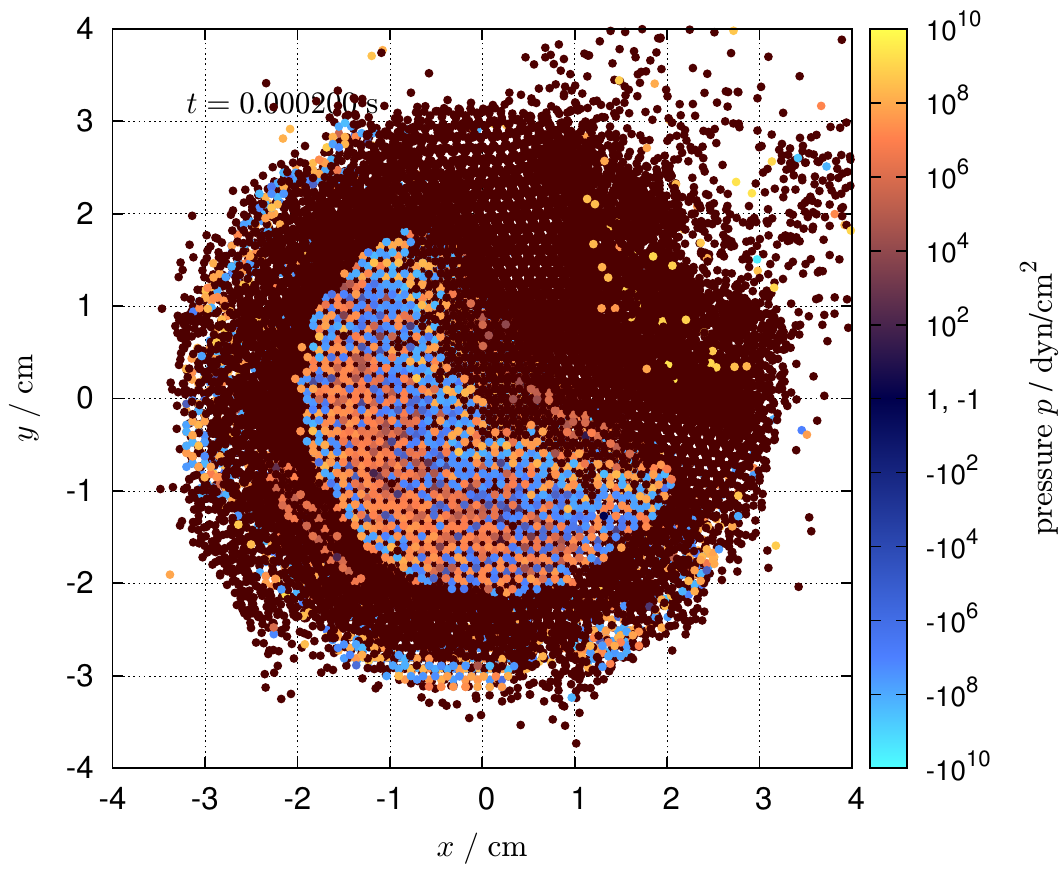}
\caption{A simulation of the classical Nakamura (1993) experiment,
but prolonged up to $200\,\mu{\rm s}$, which exhibits problems
with energy conservation, as explained in the main text.
We show a~cross section in the $(x, y)$ plane and pressure~$P$
in colour logarithmic scale. There are acoustic waves with wavelengths
close to the resolution limit in the inner monolithic cavity,
    surrounded by fully damaged material (with ${\cal D} = 1$).
In our setup,
$D_{\rm target} = 6\,{\rm cm}$,
$d_{\rm project} = 0.7\,{\rm cm}$,
$\rho = 2.7$, or
$1.15\,{\rm g}\,{\rm cm}^{-3}$ respectively,
$v_{\rm imp} = 3.2\,{\rm km}\,{\rm s}^{-1}$,
$\phi_{\rm imp} = 30^\circ$,
$N_{\rm part} \doteq 7\cdot10^5$.}
\label{sph_Nakamura_1993_NPART50000_xy_PRESSURE2_00199}
\end{figure}

Even though we always start with intact monolithic targets,
we realized that prolonged computations of the fragmentation phase
require a more careful treatment of undamaged/damaged boundaries.
The reason is the following rather complicated mechanism:
(i)~The shock wave, followed by a decompression wave, {\em partially\/}
destroys the target. After the reflection from the free surface,
the rarefaction (or sound) wave propagates back to the target.
(ii)~However, neither wave can propagate into already damaged parts,
so there is only an undamaged cavity.
(iii)~This cavity has an {\em irregular\/} boundary, so that reflections
from it create a lot of small waves, interfering with each other.
(iv)~As a result of this interference, there is a lot of particles
that have either high positive or high negative pressure,
so that the pressure gradient --- computed as a sum over neighbours --- is {\em zero\/}!
(v)~$\nabla P = 0$ means no motion, and consequently no pressure release
is possible.
(vi)~However, at the {\em boundary\/} between undamaged/damaged material,
there are some particles with $P > 0$,
next to the damaged ones with $P = 0$,
which slowly push away the undamaged particles in the surroundings.
(vii)~Because the pressure is still not released, the steady pushing
eventually destroys the whole target
(see Fig.~\ref{sph_Nakamura_1993_NPART50000_xy_PRESSURE2_00199}).

In reality, this does not happen, because the waves can indeed
become very small and dissipate. In SPH, the dissipation
of waves at the resolution limit is impossible.
Increasing resolution does not help at all ---
the boundary is even more irregular and the sound waves
will anyway become as small as the resolution.

As a solution, we can use an upper limit for damage, very close to~1,
but not equal to 1, e.g. $(1-{\cal D}) = 10^{-12}$. Then the acoustic waves
are damped (in a few seconds for $D = 1\,{\rm km}$ targets)
and the energy is conserved perfectly. Another option would be to use
a more detailed rheology of the material, namely the internal friction
and Drucker--Prager yield criterion \citep[as in][]{Jutzi_2015}.

\section{Additional figures}
Figures D.\,10 to D.\,21 show the situation for non-standard impact angles.

\begin{figure*}
\centering
\hbox{%
    \hskip-70pt%
    \includegraphics[width=17.5cm]{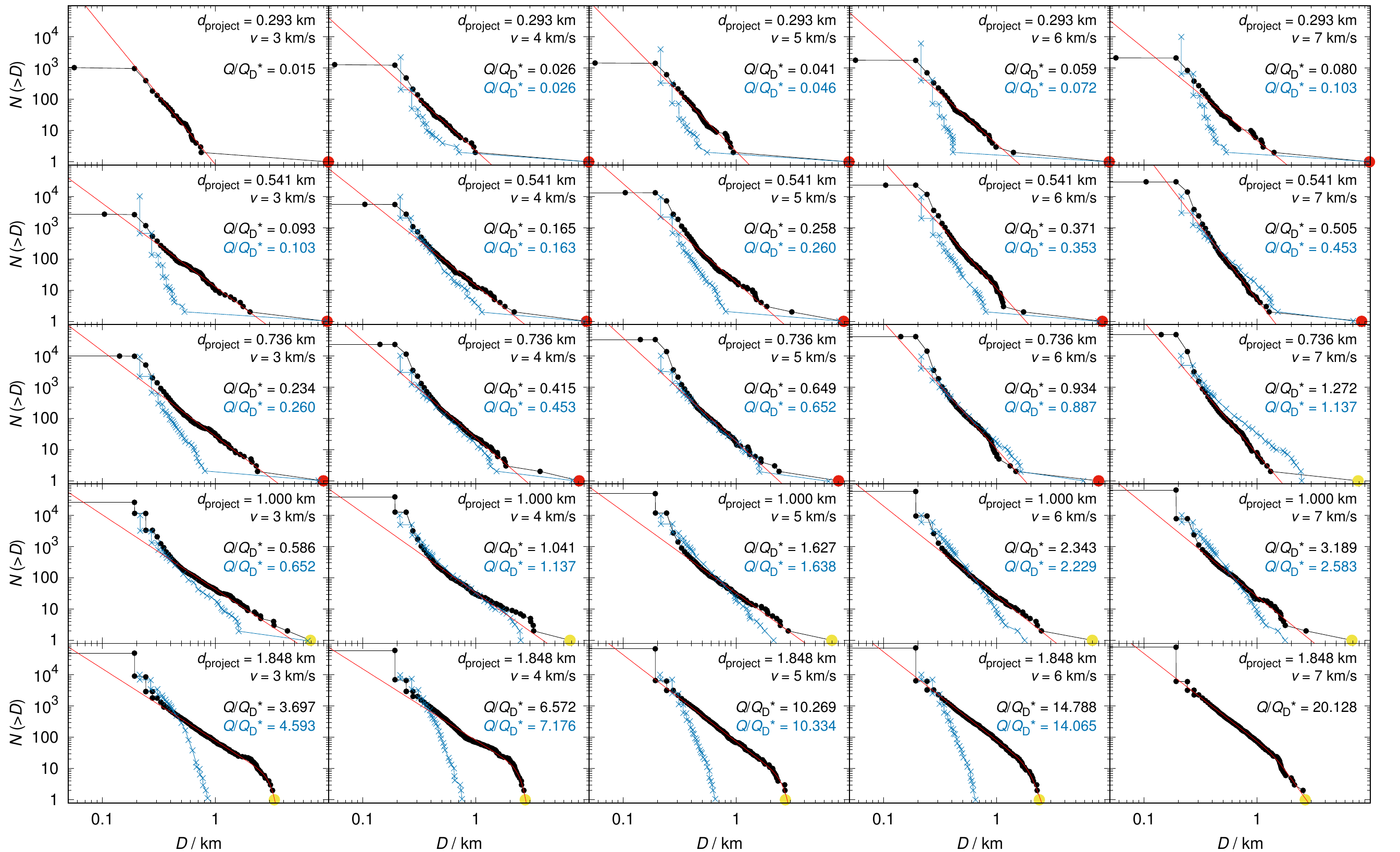}
    }
    \caption{Impact angle 15$^\circ$.}
    \label{GRID_1km_size_distribution_SCALED_100km}
\end{figure*}
\begin{figure*}
\centering
\hbox{%
    \hskip-70pt%
    \includegraphics[width=17.5cm]{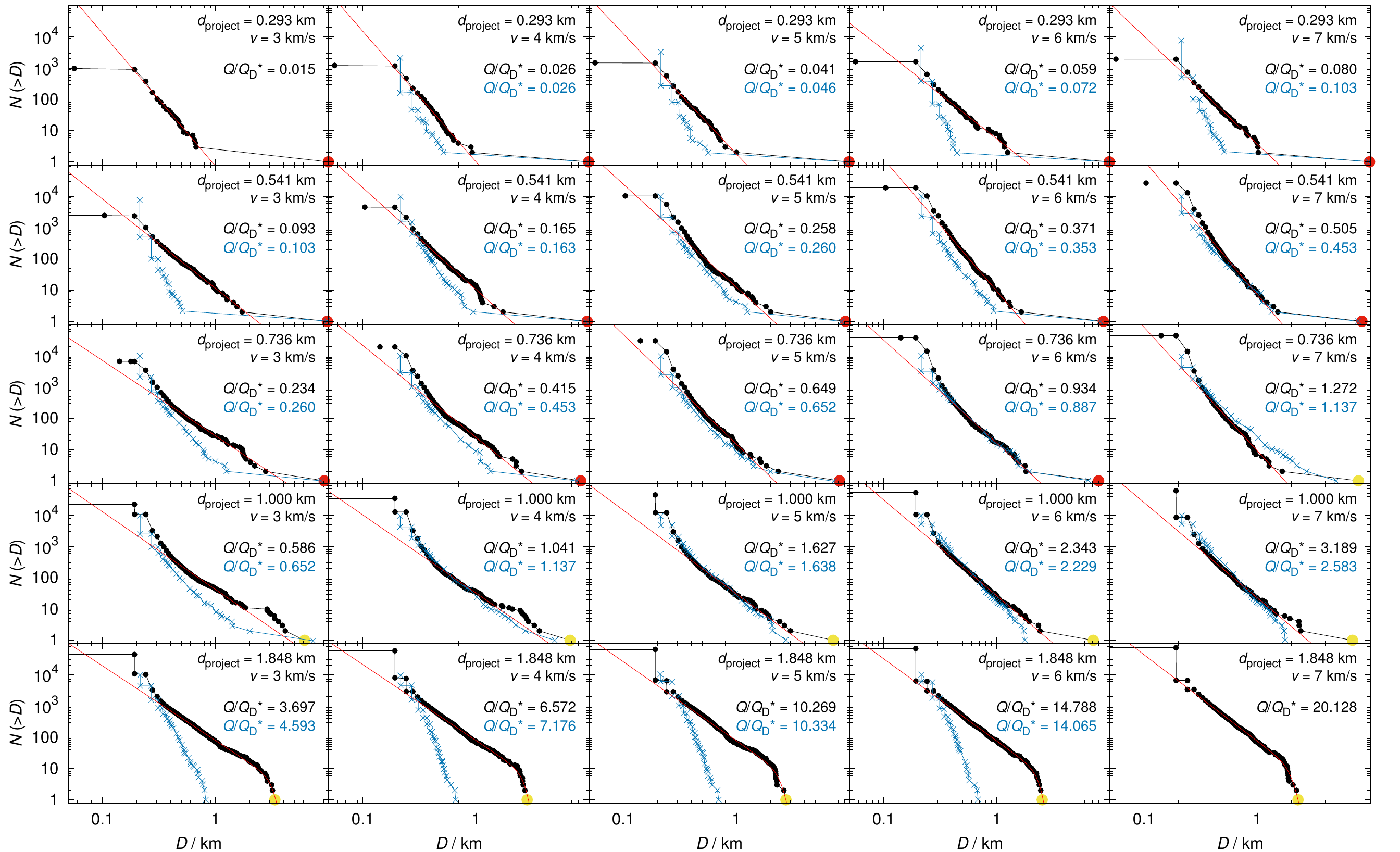}
    }
    \caption{Impact angle 30$^\circ$.}%
\end{figure*}
\begin{figure*}
\centering
\hbox{%
    \hskip-70pt%
    \includegraphics[width=17.5cm]{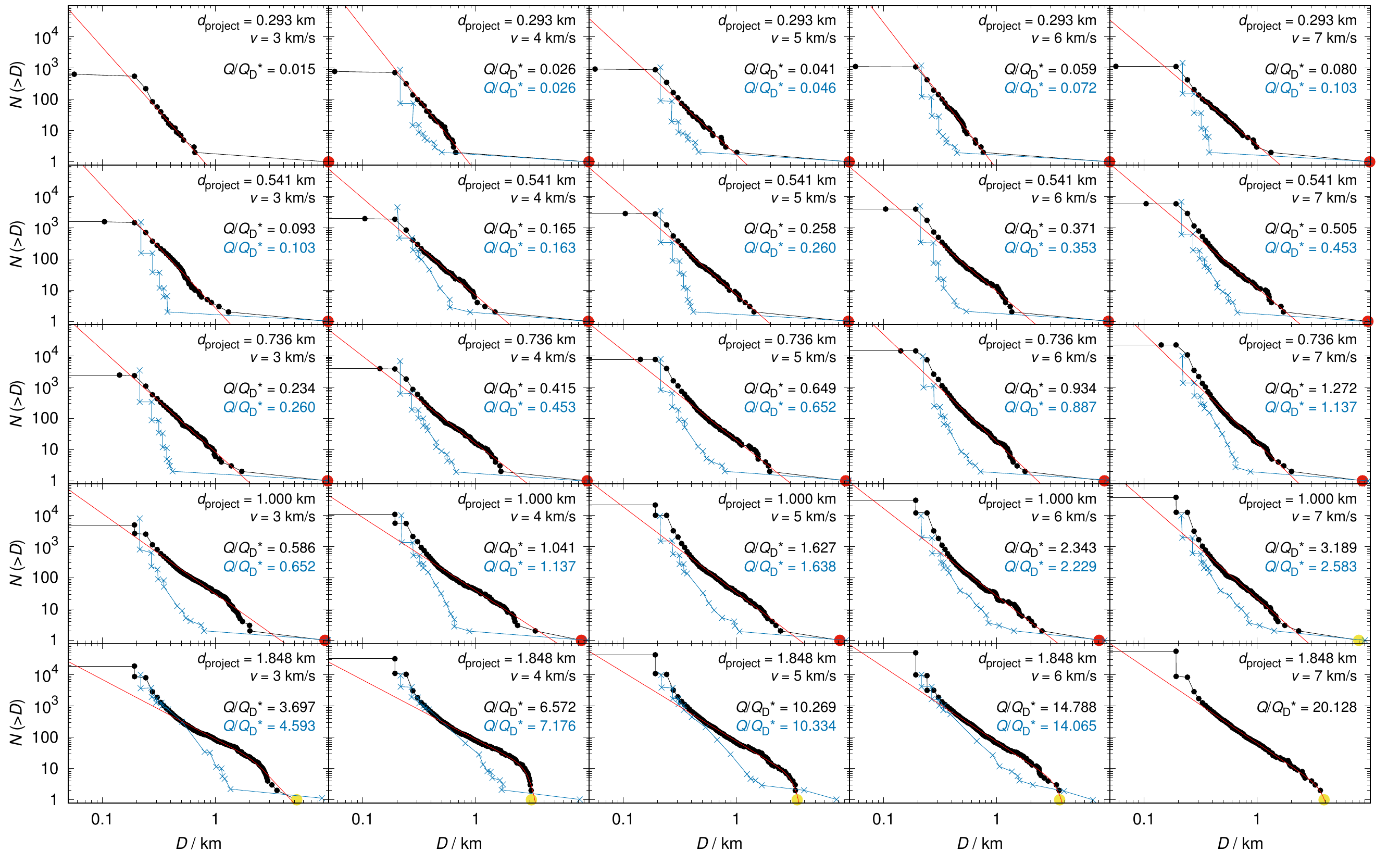}
    }
\caption{Impact angle 60$^\circ$.}
\end{figure*}
\begin{figure*}
\centering
    \hbox{%
    \hskip-70pt%
    \includegraphics[width=17.5cm]{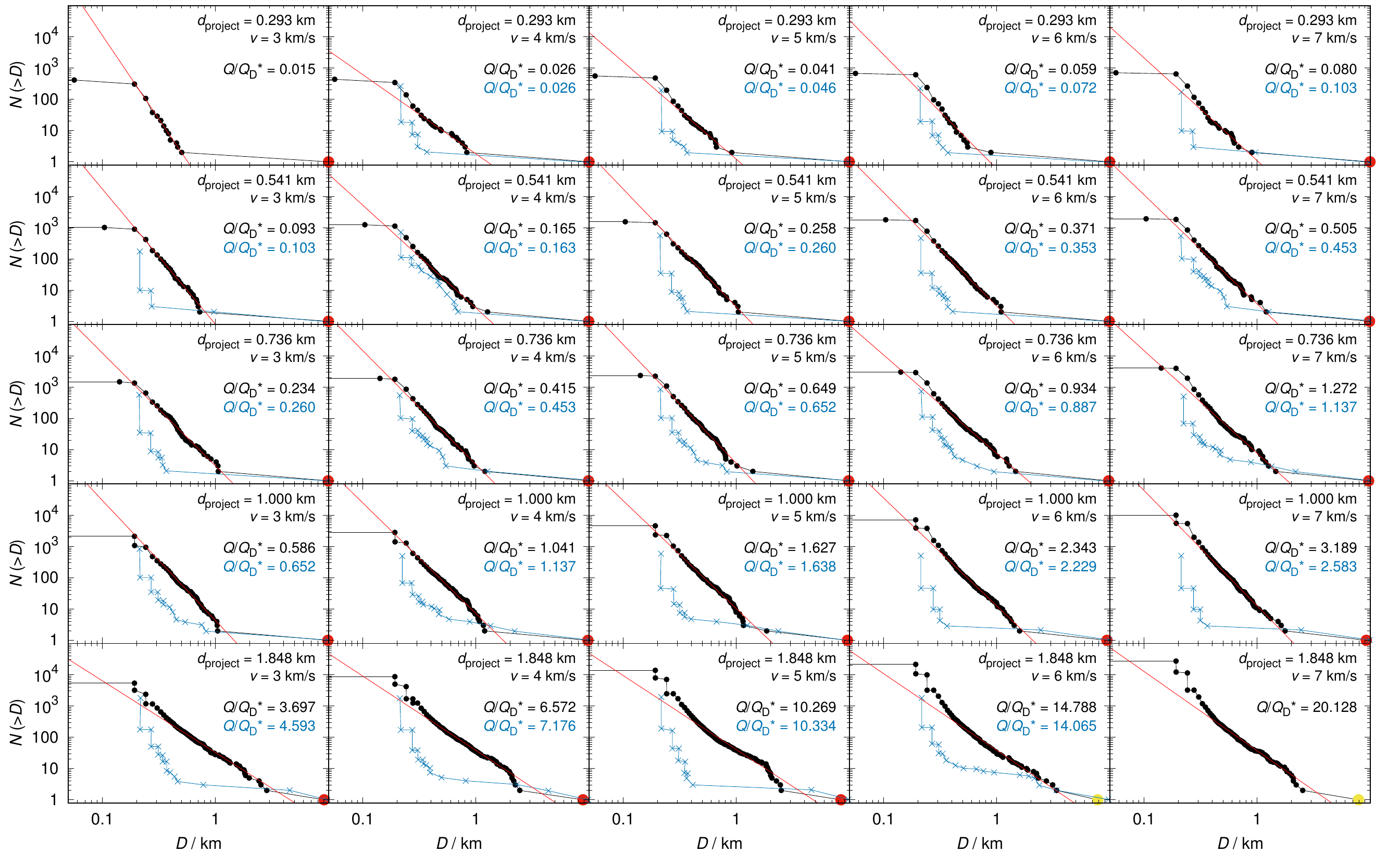}
    }
\caption{Impact angle 75$^\circ$.}
\end{figure*}

\begin{figure*}
\centering
\hbox{%
    \hskip-70pt%
\includegraphics[width=18cm]{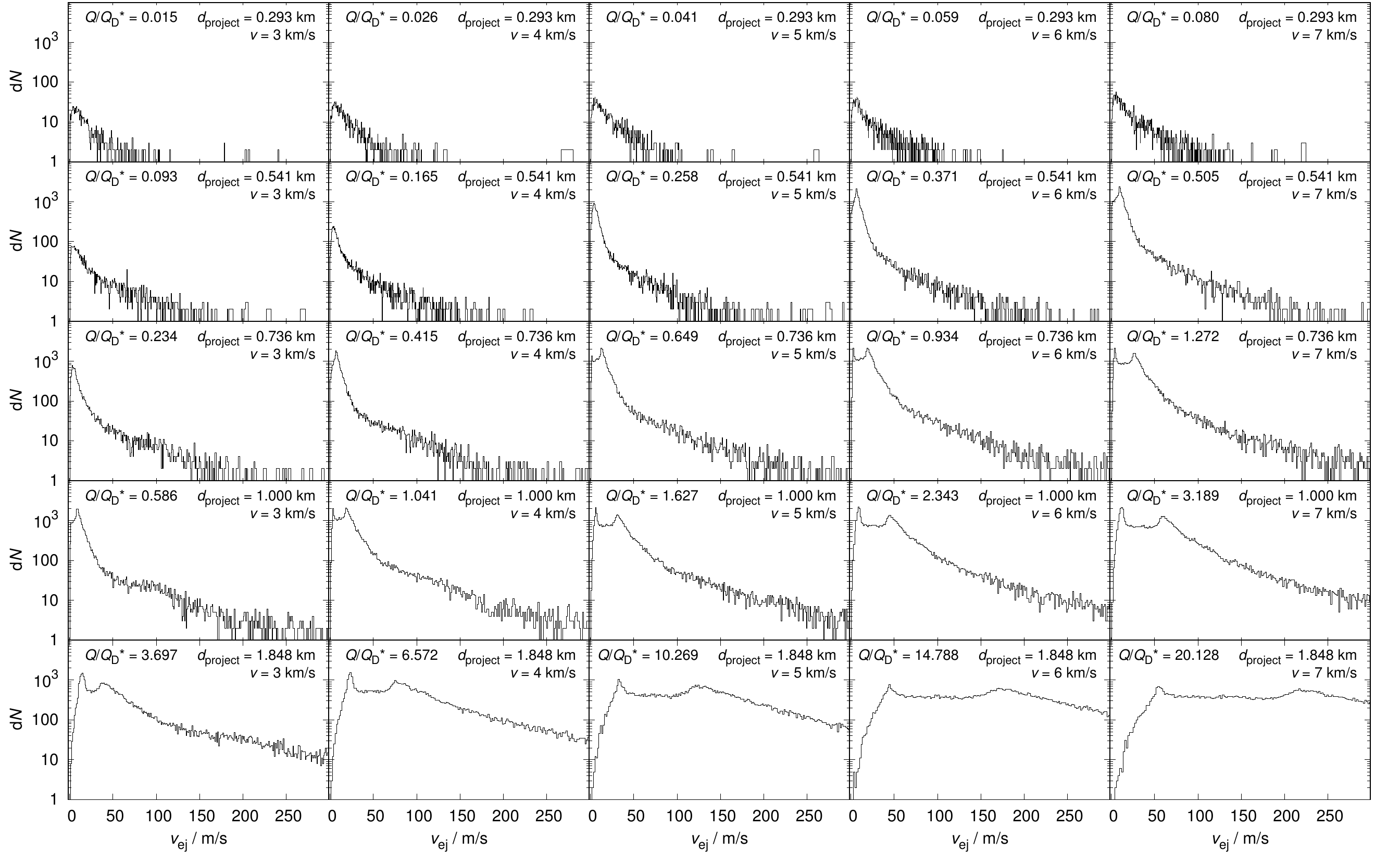}
    }
\caption{Impact angle 15$^\circ$.}
\end{figure*}
\begin{figure*}
\centering
\hbox{%
    \hskip-70pt%
\includegraphics[width=18cm]{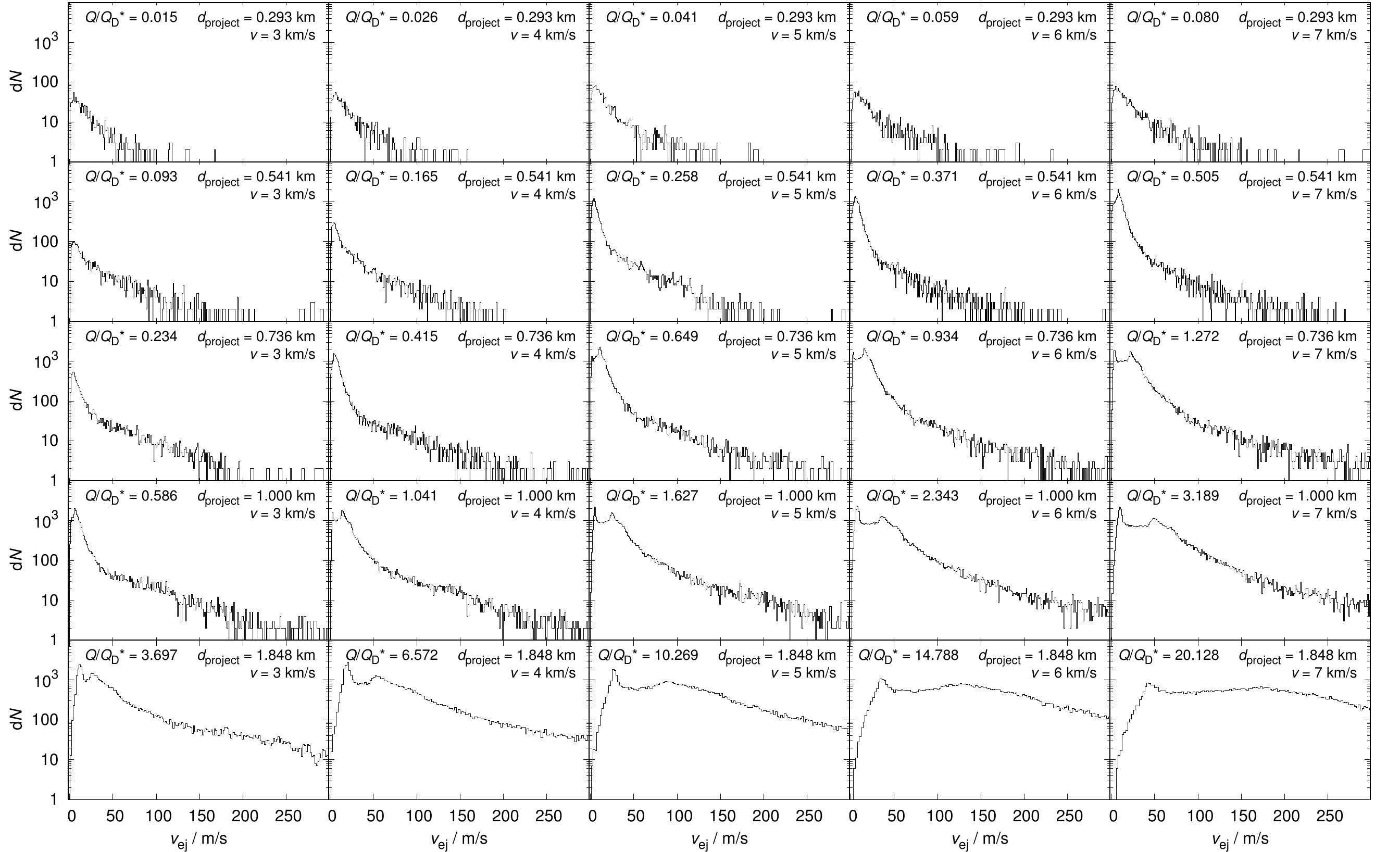}
    }
\caption{Impact angle 30$^\circ$.}
\end{figure*}
\begin{figure*}
\centering
\hbox{%
    \hskip-70pt%
\includegraphics[width=18cm]{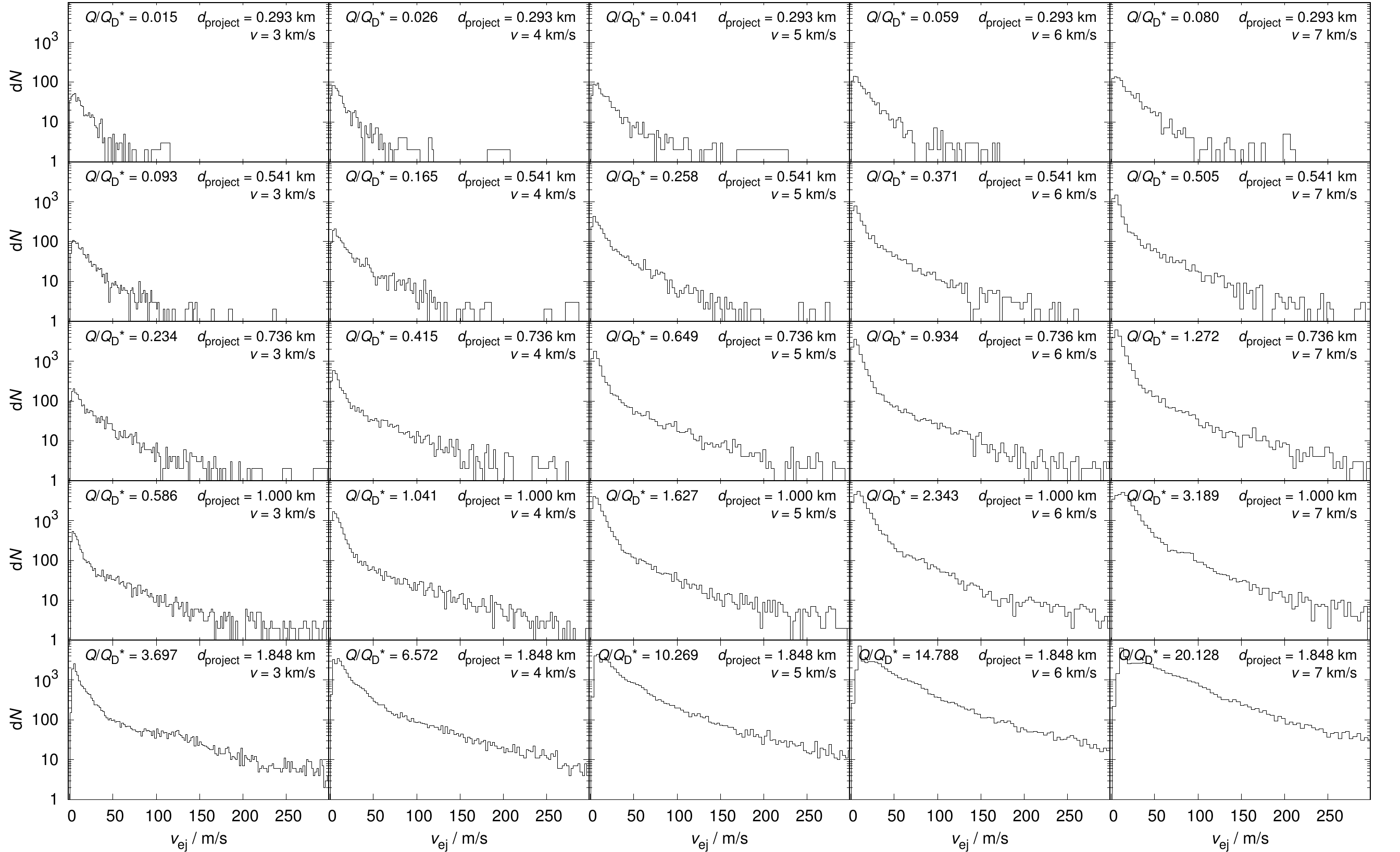}
    }
\caption{Impact angle 60$^\circ$.}
\end{figure*}
\begin{figure*}
\centering
\hbox{%
    \hskip-70pt%
\includegraphics[width=18cm]{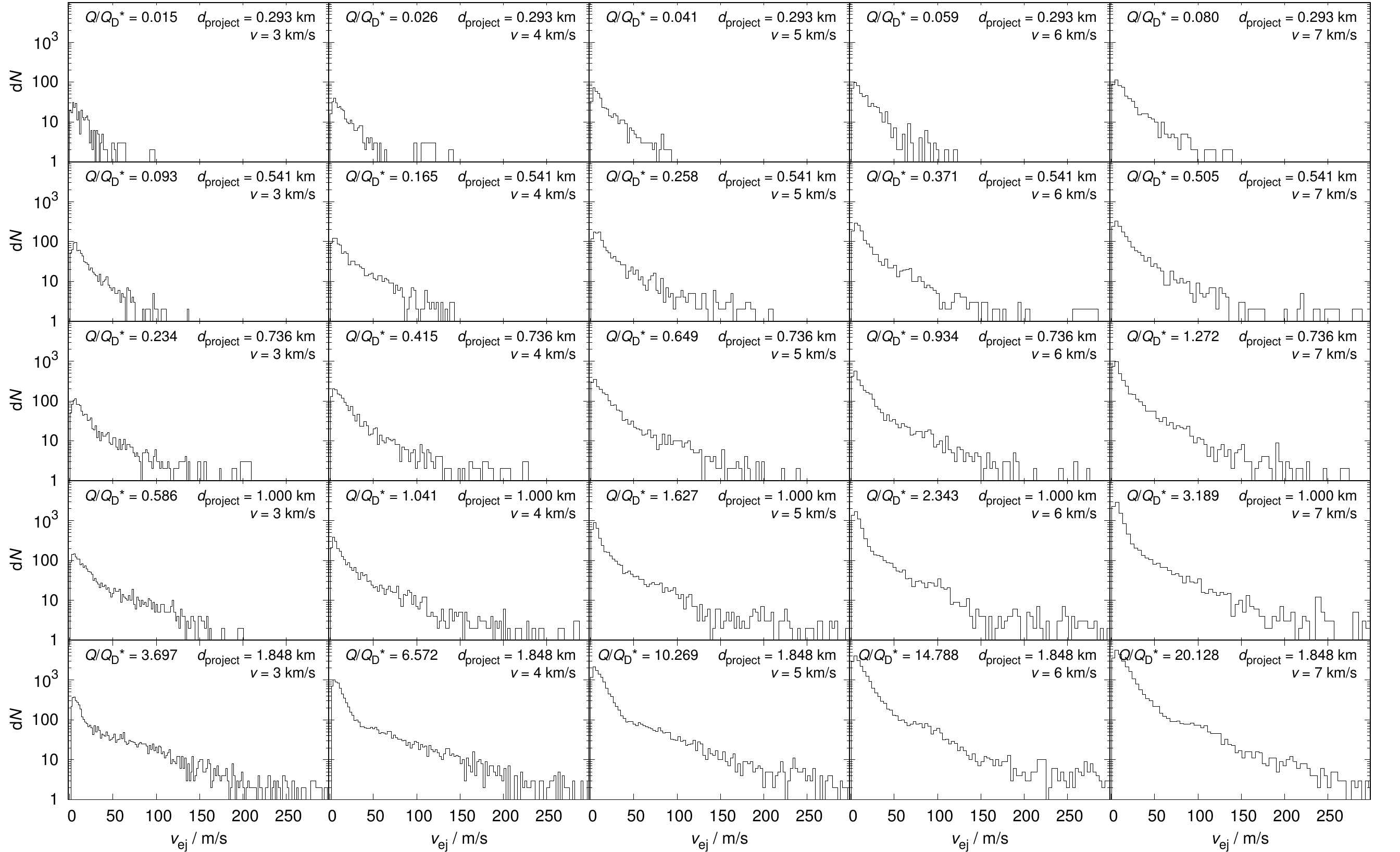}
    }
\caption{Impact angle 75$^\circ$.}
\end{figure*}

\begin{figure*}
\centering
\hbox{%
    \hskip-70pt%
\includegraphics[width=18cm]{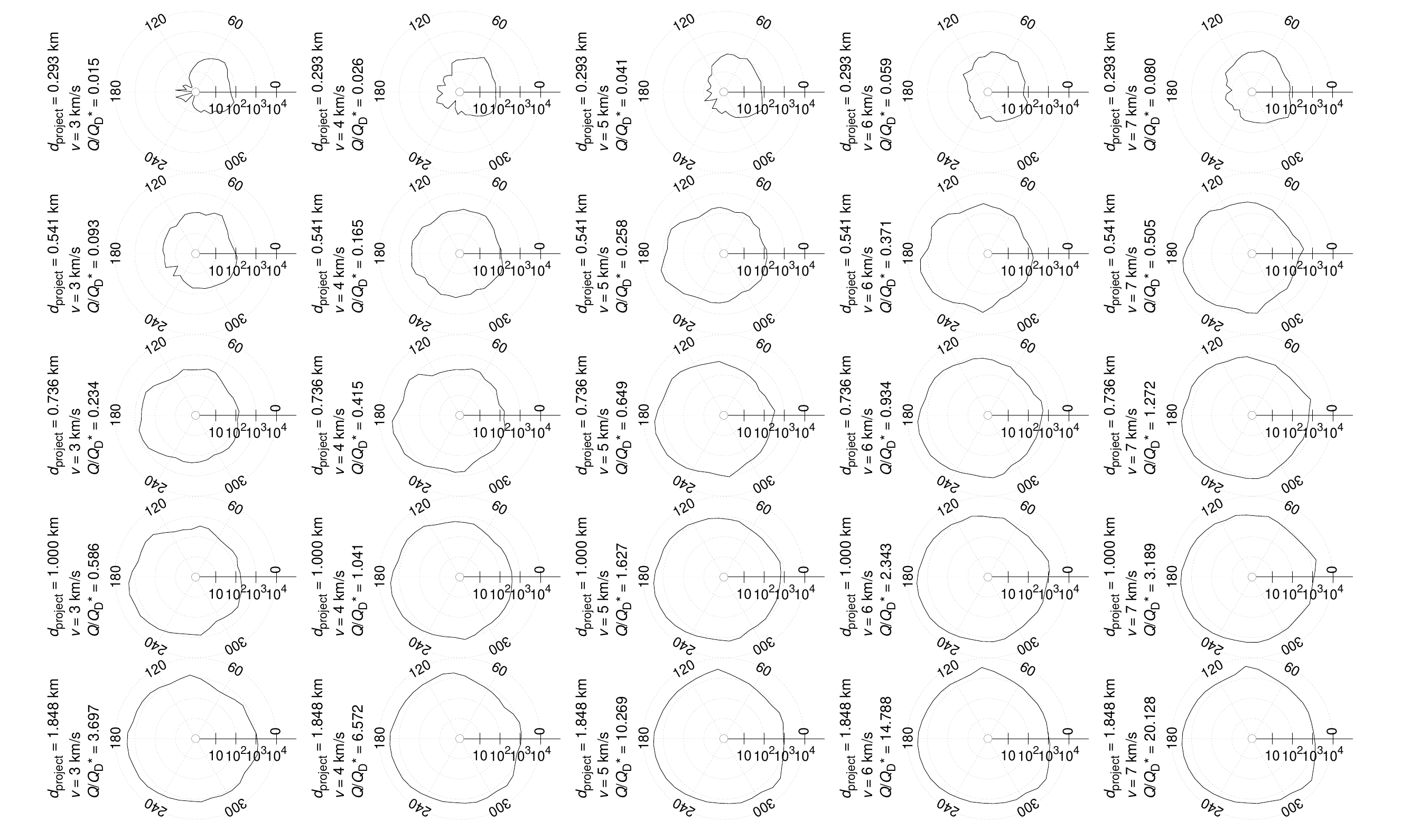}
    }
\caption{Impact angle 15$^\circ$.}
\end{figure*}
\begin{figure*}
\centering
\hbox{%
    \hskip-70pt%
\includegraphics[width=18cm]{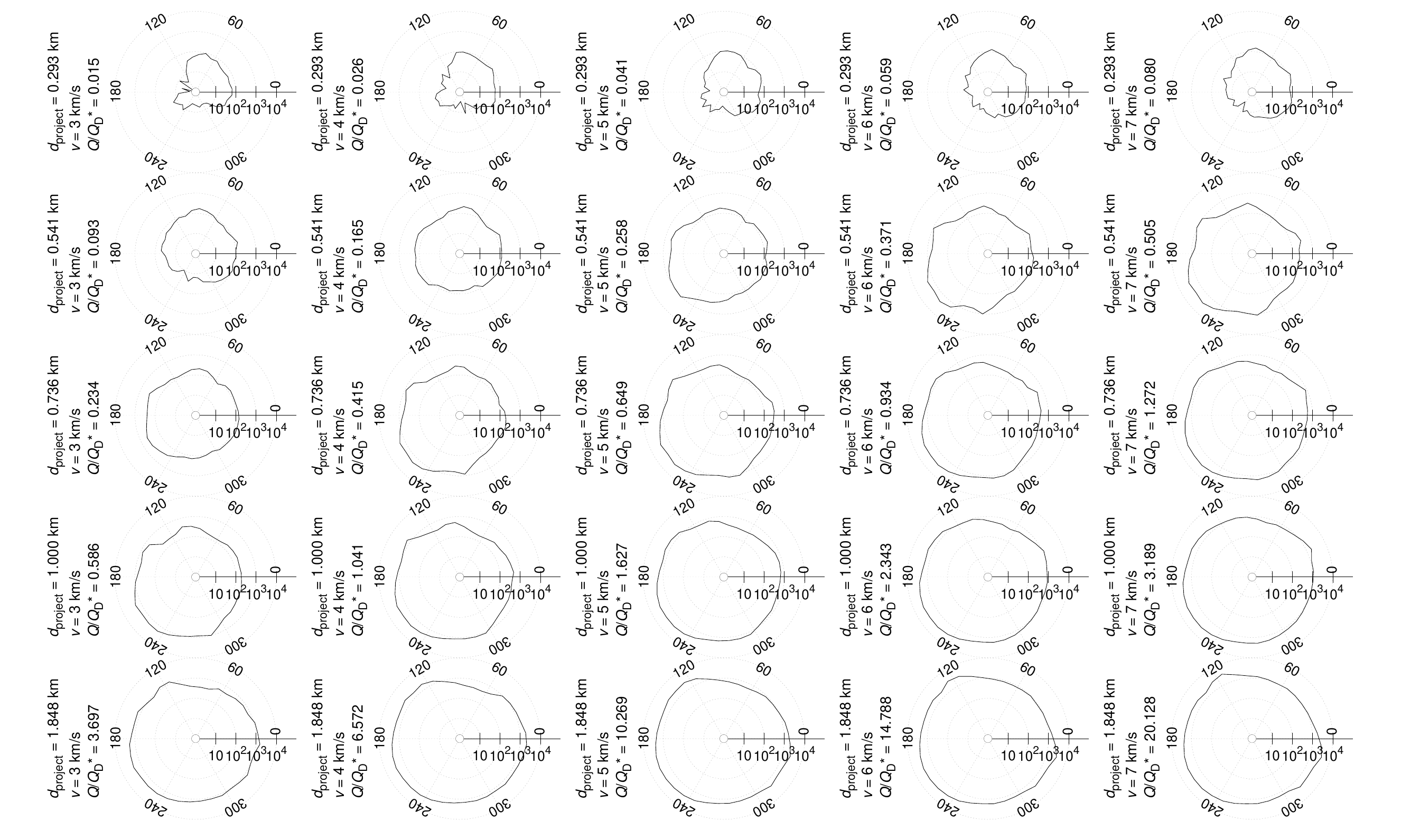}
    }
\caption{Impact angle 30$^\circ$.}
\end{figure*}
\begin{figure*}
\centering
\hbox{%
    \hskip-70pt%
\includegraphics[width=18cm]{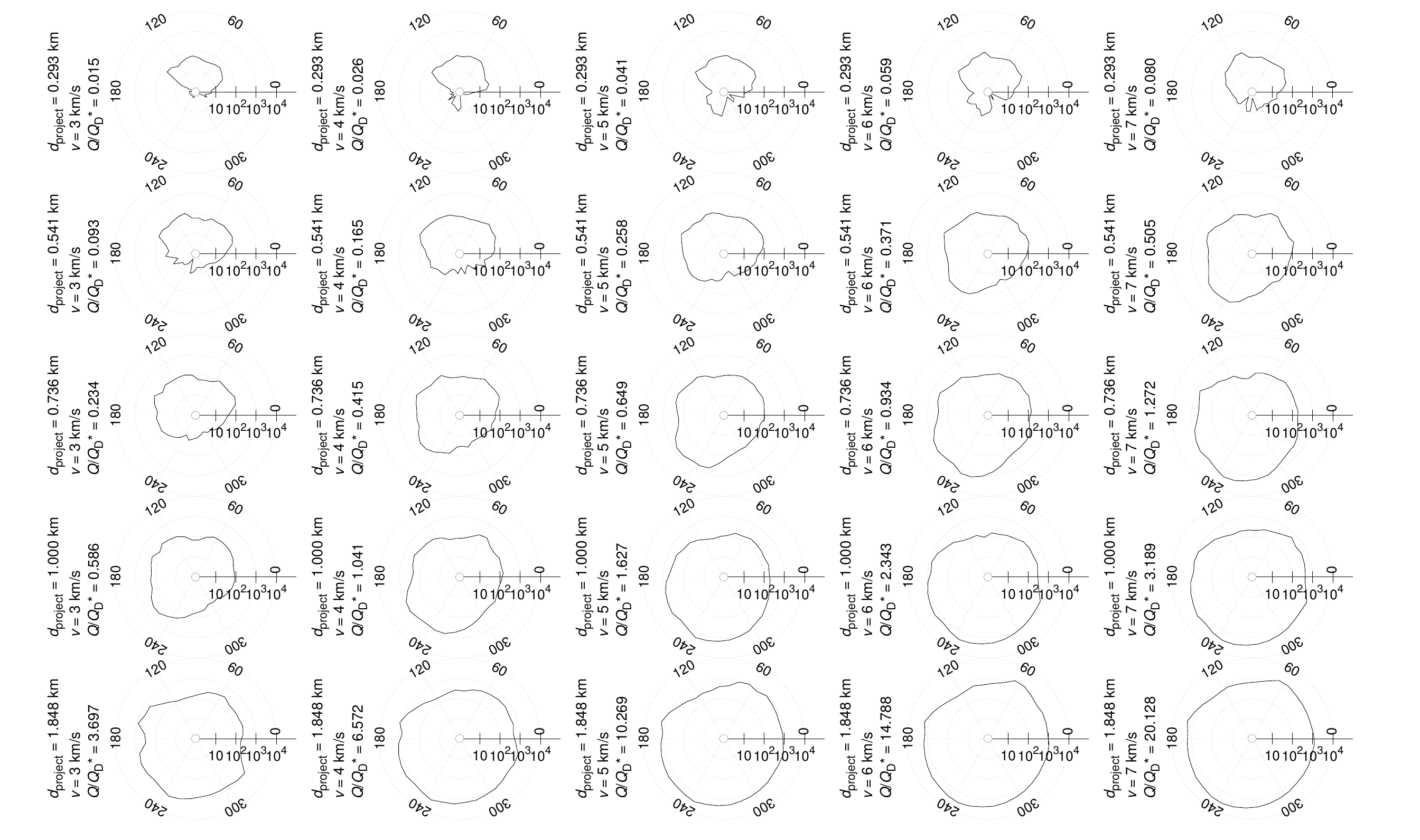}
    }
\caption{Impact angle 60$^\circ$.}
\end{figure*}
\begin{figure*}
\centering
\hbox{%
    \hskip-70pt%
\includegraphics[width=18cm]{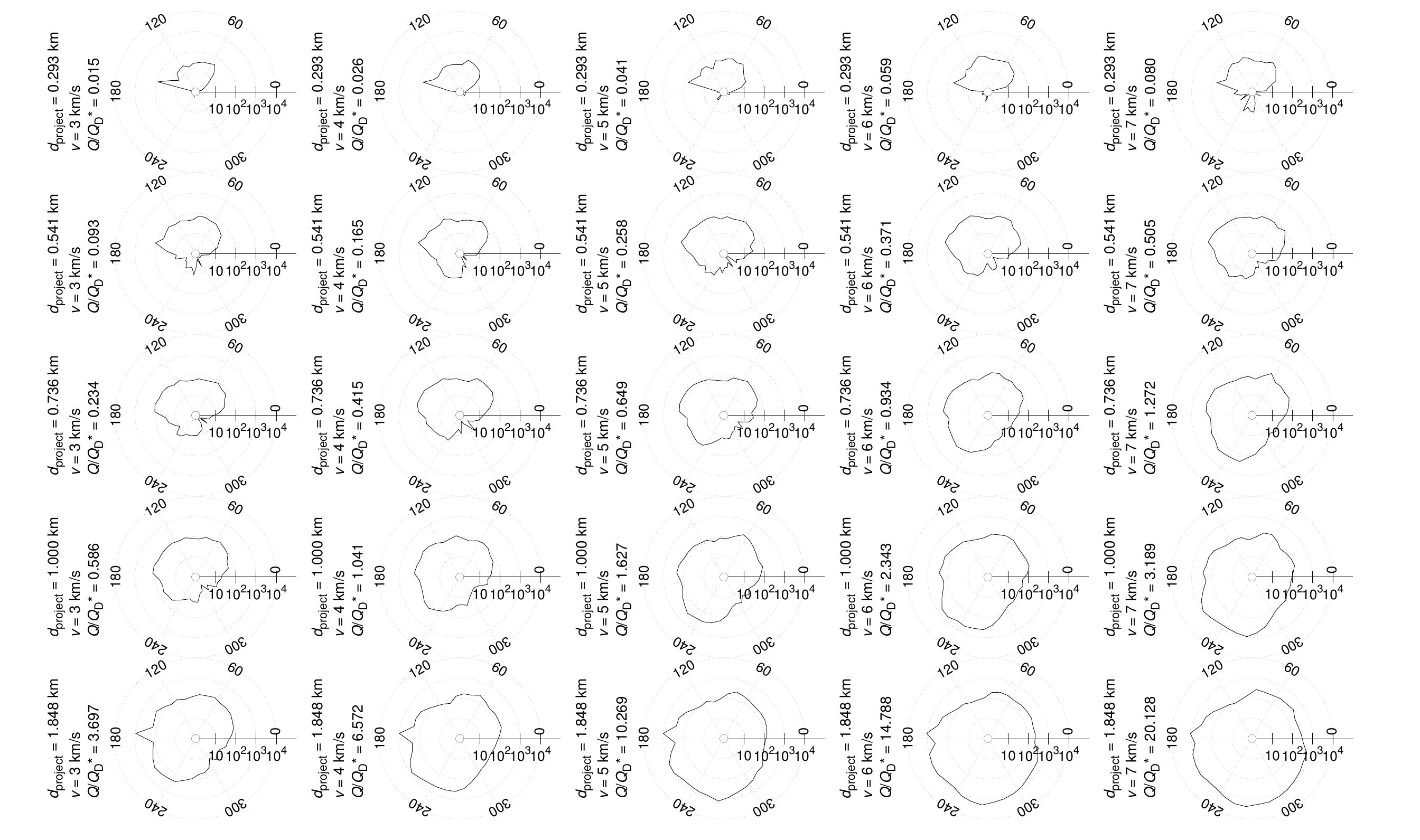}
    }
\caption{Impact angle 75$^\circ$.}
\end{figure*}


\bibliographystyle{elsarticle-harv}

\end{document}